\documentclass[journal]{IEEEtran}
\usepackage{amsmath,amsfonts}
\usepackage{array}
\usepackage[caption=false,font=normalsize,labelfont=sf,textfont=sf]{subfig}
\usepackage{textcomp}
\usepackage{stfloats}
\usepackage{url}
\usepackage{verbatim}
\usepackage{graphicx}
\usepackage{cite}
\usepackage{balance}
\usepackage[inline]{enumitem}
\usepackage{nicematrix}
\usepackage[toc,acronym]{glossaries}
\usepackage{hyperref}
\usepackage{orcidlink}
\usepackage{booktabs}
\usepackage{algorithmic}
\usepackage[ruled,vlined]{algorithm2e}
\usepackage{pifont}%
\newcommand{\cmark}{\ding{51}}%
\newcommand{\xmark}{\ding{55}}%
\newcommand{\highlight}[1]{\textcolor{black}{#1}}
\hypersetup{
    colorlinks=true,
    linkcolor=blue,
    filecolor=magenta,
    urlcolor=cyan,
    }
\newacronym{6lowpan}{6LoWPAN}{IPv6 over Low-Power Wireless Personal Area Networks}
\newacronym{ai}{AI}{Artificial Intelligence}
\newacronym{ap}{AP}{Access Point}
\newacronym{api}{API}{Application Program Interface}
\newacronym{ae}{AE}{Autoencoder}
\newacronym{automl}{AutoML}{Automated Machine Learning}
\newacronym{asn}{ASN}{Absolute Slot Number}
\newacronym{beassa}{BEA-SSA}{Bald Eagle Assisted SSA}
\newacronym{bs}{BS}{Base Station}
\newacronym{bfs}{BFS}{Breadth First Search}
\newacronym{blip}{BLIP}{Berkeley Low-power IP}
\newacronym{ch}{CH}{Cluster Head}
\newacronym{ctp}{CTP}{Collection Tree Protocol}
\newacronym{cnn}{CNN}{Convolutional Neural Networks}
\newacronym{dt}{DT}{Decision Tree}
\newacronym{dcnn}{DCNN}{Deep Convolutional Neural Network}
\newacronym{dl}{DL}{Deep Learning}
\newacronym{dqn}{DQN}{Deep Q-Learning}
\newacronym{dodag}{DODAG}{Destination Oriented Directed Acyclic Graph}
\newacronym{drl}{DRL}{Deep Reinforcement Learning}
\newacronym{eos}{EOS}{Embedded Operating System}
\newacronym{eb}{EB}{Enhanced Beacon}
\newacronym{ewma}{EWMA}{Exponential Weighted Moving Average}
\newacronym{fnd}{FND}{First Node Death}
\newacronym{fpga}{FPGA}{Field-Programmable Gate Array}
\newacronym{fsm}{FSM}{Finite State Machine}
\newacronym{gan}{GAN}{Generative Adversarial Network}
\newacronym{gprs}{GPRS}{General Packet Radio Service}
\newacronym{gps}{GPS}{Global Positioning System}
\newacronym{glpk}{GLPK}{GNU Linear Programming Kit}
\newacronym{hnd}{HND}{Half Node Death}
\newacronym{hrl}{HRL}{Hierarchical Reinforcement Learning}
\newacronym{hrltsch}{HRL-TSCH}{Hierarchical Reinforcement Learning-based Time Slotted Channel Hopping}
\newacronym{ia}{IA}{Intelligent Agent}
\newacronym{icmpv6}{ICMPv6}{Internet Control Message Protocol version 6}
\newacronym{ietf}{IETF}{Internet Engineering Task Force}
\newacronym{iot}{IoT}{Internet of Things}
\newacronym{iiot}{IIoT}{Industrial Internet of Things}
\newacronym{ids}{IDS}{Intrusion Detection System}
\newacronym{ip}{IP}{Internet Protocol}
\newacronym{ipv4}{IPv4}{Internet Protocol version 4}
\newacronym{ipv6}{IPv6}{Internet Protocol version 6}
\newacronym{knn}{k-NN}{K-Nearest Neighbour}
\newacronym{kpi}{KPI}{Key Performance Indicator}
\newacronym{lan}{LAN}{Local Area Network}
\newacronym{leach}{LEACH}{Low-Energy Adaptive Clustering Hierarchy}
\newacronym{leach-c}{LEACH-C}{Low-Energy Adaptive Clustering Hierarchy with Centralized Controller}
\newacronym{leach-rlc}{LEACH-RLC}{Low-Energy Adaptive Clustering Hierarchy with Reinforcement Learning-based Controller}
\newacronym{wlan}{WLAN}{Wireless Local Area Network}
\newacronym{lora}{LoRa}{Long Range}
\newacronym{lorawan}{LoRaWAN}{Long Range Wide Area Network}
\newacronym{lowpan}{LoWPAN}{Low-Power Wireless Personal Area Networks}
\newacronym{lqi}{LQI}{Link Quality Indicator}
\newacronym{m2m}{M2M}{Machine-to-Machine}
\newacronym{mac}{MAC}{Media Access Control}
\newacronym{mdp}{MDP}{Markov Decision Process}
\newacronym{mems}{MEMS}{Micro-Electro-Mechanical Systems}
\newacronym{mcu}{MCU}{Microcontroller Unit}
\newacronym{milp}{MILP}{Mixed Integer Linear Programming}
\newacronym{ml}{ML}{Machine Learning}
\newacronym{mlsdwsn}{ML-SDWSN}{Machine Learning Software-Defined Wireless Sensor Network}
\newacronym{mst}{MST}{Minimum Spanning Tree}
\newacronym{na}{NA}{Neighbor Advertisement}
\newacronym{nc}{NC}{Network Configuration}
\newacronym{nes}{NES}{Networked Embedded Systems}
\newacronym{nd}{ND}{Neighbor Discovery}
\newacronym{nl}{NL}{Network Lifetime}
\newacronym{nn}{NN}{Neural Network}
\newacronym{nch}{NCH}{Non-Cluster Head}
\newacronym{os}{OS}{Operating System}
\newacronym{pdr}{PDR}{Packet Delivery Ratio}
\newacronym{plr}{PLR}{Packet Loss Rate}
\newacronym{pso}{PSO}{Particle Swarm Optimisation}
\newacronym{pca}{PCA}{Principal Component Analysis}
\newacronym{qos}{QoS}{Quality of Service}
\newacronym{ram}{RAM}{Random-Access Memory}
\newacronym{rdc}{RDC}{Radio Duty-Cycle}
\newacronym{rom}{ROM}{Read-Only Memory}
\newacronym{rnn}{RNN}{Recurrent Neural Network}
\newacronym{rl}{RL}{Reinforcement Learning}
\newacronym{rtt}{RTT}{Round-Trip Time}
\newacronym{rpl}{RPL}{Routing Protocol for Low-Power and Lossy Networks}
\newacronym{rssi}{RSSI}{Received Signal Strength Indicator}
\newacronym{stp}{STP}{Spanning Tree Protocol}
\newacronym{svm}{SVM}{Support Vector Machine}
\newacronym{sdn}{SDN}{Software-Defined Networking}
\newacronym{snr}{SNR}{Signal to Noise Ratio}
\newacronym{slip}{SLIP}{Serial Line Internet Protocol}
\newacronym{sdwsn}{SDWSN}{Software-Defined Wireless Sensor Network}
\newacronym{soh}{SOH}{State of Health}
\newacronym{sp}{SP}{Shortest Path}
\newacronym{tdma}{TDMA}{Time Division Multiple Access}
\newacronym{tl}{TL}{Transfer Learning}
\newacronym{tlm}{TLM}{Traffic Load Minimisation}
\newacronym{tcp}{TCP}{Transmission Control Protocol}
\newacronym{tsch}{TSCH}{Time Slotted Channel Hopping}
\newacronym{udp}{UDP}{User Datagram Protocol}
\newacronym{uip}{$\mu$IP}{micro Internet Protocol}
\newacronym{uipv6}{$\mu$IPv6}{micro Internet Protocol version 6}
\newacronym{wam}{WAM}{Weighted Arithmetic Mean}
\newacronym{wban}{WBAN}{Wireless Body Area Network}
\newacronym{wpan}{WPAN}{Wireless Personal Area Network}
\newacronym{wsan}{WSAN}{Wireless Sensor and Actuator Network}
\newacronym{wsn}{WSN}{Wireless Sensor Network}

\begin{document}

\title{LEACH-RLC: Enhancing IoT Data Transmission with Optimized Clustering and Reinforcement Learning}

\author{F. Fernando~Jurado-Lasso\,\orcidlink{0000-0002-5005-781X}, \IEEEmembership{Member, IEEE},
  J. F. Jurado\,\orcidlink{0000-0001-5193-8566},
  and Xenofon~Fafoutis\,\orcidlink{0000-0002-9871-0013}, \IEEEmembership{Senior Member, IEEE}
  \thanks{Manuscript received February 1, 2024; revised xx, xx. This work was partly supported by DAIS. DAIS (https://dais-project.eu/)  has received funding from the  ECSEL  Joint  Undertaking  (JU) under grant agreement No 101007273. The JU receives support from the European Union’s Horizon  2020  research and innovation programme and Sweden, Spain, Portugal, Belgium, Germany, Slovenia, Czech Republic, Netherlands, Denmark, Norway, and Turkey. The document reflects only the authors' view, and the Commission is not responsible for any use that may be made of the information it contains. Danish participants are supported by Innovation Fund Denmark under grant agreement No. 0228-00004A.}%
  \thanks{F. Fernando Jurado-Lasso and Xenofon Fafoutis are with the
    Embedded Systems Engineering section, DTU Compute, Technical University
    of Denmark, 2800 Lyngby, Denmark (e-mail: ffjla@dtu.dk; xefa@dtu.dk).}
  \thanks{J. F. Jurado is with the Department of Basic Science, Faculty of Engineering and Administration, Universidad Nacional de Colombia Sede Palmira, Palmira 763531, Colombia (e-mail: jfjurado@unal.edu.co).}
}

\markboth{Journal of \LaTeX\ Class Files,~Vol.~14, No.~8, August~2024}%
{Shell \MakeLowercase{\textit{et al.}}: A Sample Article Using IEEEtran.cls for IEEE Journals}

\maketitle

\begin{abstract}
  \acrfullpl{wsn} play a pivotal role in enabling \acrfull{iot} devices with sensing and actuation capabilities.
Operating in remote and resource-constrained environments, these \acrshort{iot} devices face challenges related to energy consumption, crucial for network longevity.
Existing clustering protocols often suffer from high control overhead, inefficient cluster formation, and poor adaptability to dynamic network conditions, leading to suboptimal data transmission and reduced network lifetime.
This paper introduces \acrfull{leach-rlc}, a novel clustering protocol designed to address these limitations by employing a \acrfull{milp} approach for strategic selection of \acrfullpl{ch} and node-to-cluster assignments.
Additionally, it integrates a \acrfull{rl} agent to minimize control overhead by learning optimal timings for generating new clusters.
\acrshort{leach-rlc} aims to balance control overhead reduction without compromising overall network performance.
Through extensive simulations, this paper investigates the frequency and opportune moments for generating new clustering solutions.
Results demonstrate the superior performance of \acrshort{leach-rlc} over state-of-the-art protocols, showcasing enhanced network lifetime, reduced average energy consumption, and minimized control overhead.
The proposed protocol contributes to advancing the efficiency and adaptability of \acrshortpl{wsn}, addressing critical challenges in \acrshort{iot} deployments.

\end{abstract}

\begin{IEEEkeywords}
  \acrfull{iot}, Clustering Protocols, \acrfull{rl}, Control Overhead, \acrfull{leach}
\end{IEEEkeywords}

\section{Introduction}
\IEEEPARstart{T}he \acrfull{iot} refers to the interconnection of devices, sensors, and actuators to the Internet, \highlight{playing a key role in modern applications such as smart cities, smart homes, and industrial automation}~\cite{atzoriInternetThingsSurvey2010,schwabFourthIndustrialRevolution2024}.
\highlight{An essential component of the \acrshort{iot} is the \acrfull{wsn}, which provides the infrastructure for efficient data collection and communication~\cite{qiuHowCanHeterogeneous2018}.}

\acrshortpl{wsn} are often deployed in remote areas where \highlight{battery replacement} is impractical~\cite{luoSurveyRoutingProtocols2021,caiBatteryfreeWirelessSensor2022}.
Therefore, \highlight{minimizing} energy consumption \highlight{is crucial to extending the network's lifetime}.
\highlight{Clustering-based protocols address this challenge by grouping nodes into clusters, with designated \acrfullpl{ch} responsible for aggregating and forwarding data to the \acrfull{bs}.}
\highlight{However, \acrshortpl{ch} consume more energy than cluster members, making their selection critical to network performance~\cite{battaBatteryStateofhealthPredictionbased2022}.}
\IEEEpubidadjcol
Among clustering protocols, \acrfull{leach}~\cite{heinzelmanEnergyefficientCommunicationProtocol2002} has gained popularity due to its simplicity and low overhead.
\highlight{It operates in a self-organizing manner, with nodes forming clusters and rotating roles to balance energy consumption.}
\highlight{\acrshortpl{ch}} are selected stochastically using a threshold equation, and non-\acrshort{ch} nodes join the closest cluster.
\highlight{\acrshortpl{ch} aggregate data from cluster members and transmit it to the \acrshort{bs} using a \acrfull{tdma} schedule to avoid collisions.}
\highlight{\acrshort{leach} does not guarantee even distribution of energy consumption among nodes, leading to imbalances~\cite{beheraResidualEnergybasedClusterhead2019a}.}

\begin{figure}[!t]
    \centering
    \includegraphics[width=0.5\columnwidth]{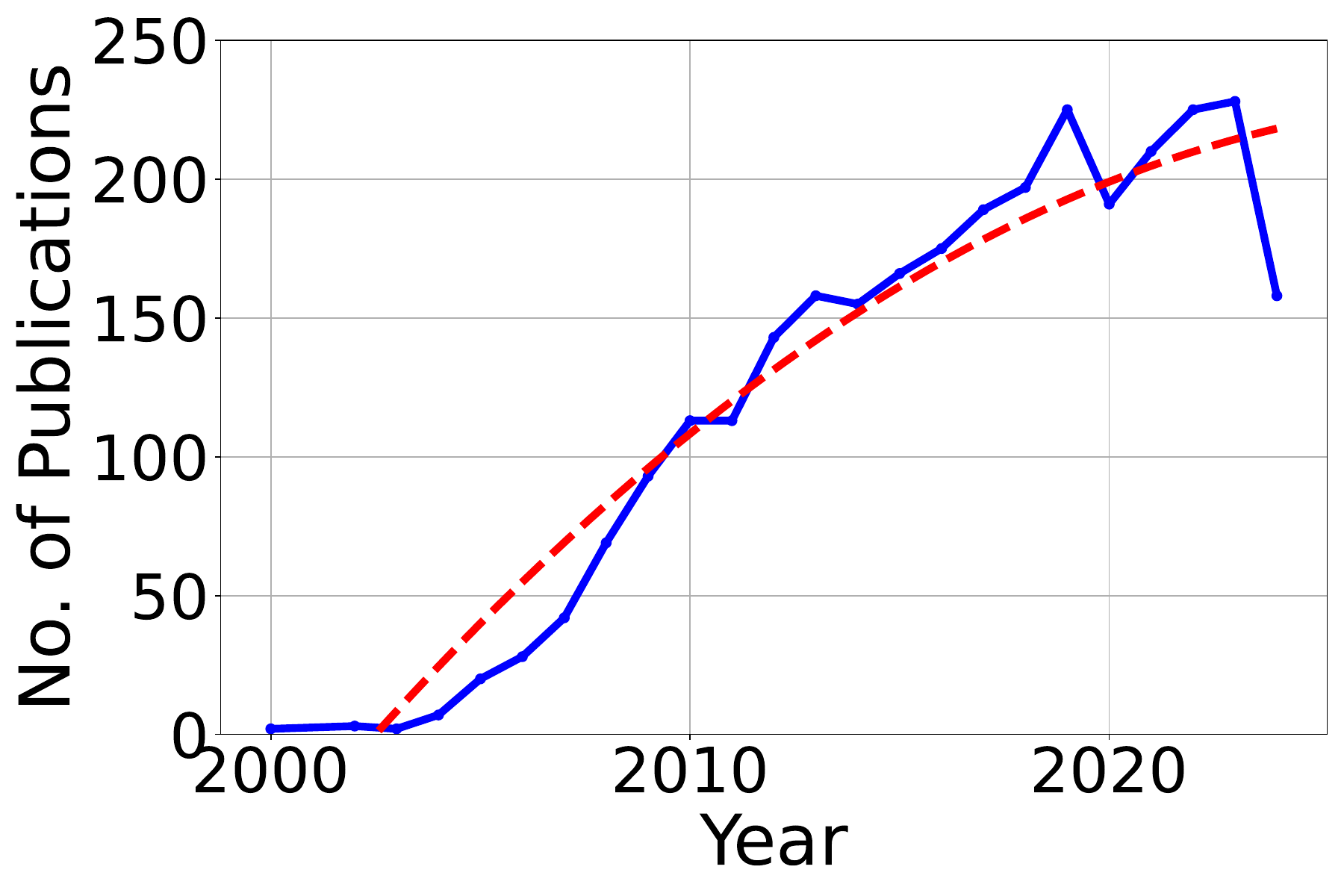}
    \caption{The number of publications on clustering protocols based on \acrshort{leach} from 2000 to 2024. The data was obtained from the Scopus database using the keywords `LEACH AND clustering AND protocol'~\cite{elsevierScopusDatabase2025}.}
    \label{fig:leach_citation_trend}
\end{figure}

\highlight{\acrfull{leach-c} was later proposed to optimize \acrshort{ch} selection by using a centralized controller~\cite{heinzelmanApplicationspecificProtocolArchitecture2002a}.}
\highlight{The controller has a global view of the network and selects \acrshortpl{ch} to minimize energy consumption.}
\highlight{despite its improvements, \acrshort{leach-c} introduces significant control overhead, lacks optimal timing for CH updates, and does not fully balance energy consumption between CHs and cluster members.}
Over the years, research on \acrshort{leach} and its variants have remained vibrant, as evidenced by the publication trends in Fig.~\ref{fig:leach_citation_trend}.

Advances in \acrshort{leach-c} clustering have improved network performance compared to \acrshort{leach}~\cite{chenRANCERandomlyCentralized2022,tebessiImprovementLEACHCProtocol2022,gamalEnhancingLifetimeWireless2022,jinCentralizedMultiHopRouting2021,maESCVADEnergySavingRouting2021}. However, these approaches are limited by several significant issues:
\begin{enumerate*}[label=(\roman*)]
    \item They rely on a centralized controller, leading to significant control overhead due to the frequent distribution of control messages.
    \item There is no optimal timing for generating new clustering solutions; sometimes maintaining the same \acrshort{ch} set for an extra round can enhance long-term performance.
    \item They focus on minimizing the squared distance between nodes and \acrshortpl{ch}, but energy consumption is also affected by the distance between \acrshortpl{ch} and the \acrshort{bs}.
    \item Finally, \acrshort{leach-c} does not balance energy consumption between \acrshortpl{ch} and cluster members, leading to imbalances.
\end{enumerate*}

Given these limitations, there is a need for more advanced clustering protocols that can dynamically adapt to network conditions, optimize energy usage, and reduce control overhead.

In this paper, we propose a novel clustering protocol, called \acrfull{leach-rlc} \footnote{The source code of \acrshort{leach-rlc} is available at \url{https://github.com/fdojurado/PyNetSim.git}}, which addresses key limitations of existing protocols such as \acrshort{leach-c}.
\acrshort{leach-rlc} combines two main innovations: (i) a \acrshort{milp} formulation for optimizing cluster formation and (ii) a reinforcement learning (RL) agent for reducing control overhead. The \acrshort{milp} optimally selects the \acrshortpl{ch} and assigns nodes to clusters to minimize the overall energy consumption, accounting for both intra-cluster (node-to-\acrshort{ch}) and inter-cluster (\acrshort{ch}-to-\acrshort{bs}) communication. Meanwhile, the RL agent dynamically determines the optimal timing to generate new clusters, avoiding unnecessary overhead caused by fixed re-clustering intervals. These features work together to improve energy efficiency, extend network lifetime, and reduce control overhead.

The key contributions of this paper are summarized as follows:

\begin{enumerate}
    \item \textbf{Introduction of LEACH-RLC:} A novel clustering protocol that incorporates a Mixed Integer Linear Programming (MILP) formulation for selecting \acrshortpl{ch} and assigning nodes to clusters. Unlike previous protocols, LEACH-RLC considers both intra-cluster energy consumption (i.e., communication between nodes and CHs) and inter-cluster energy consumption (i.e., CH-to-sink communication), ensuring a more comprehensive and energy-efficient clustering strategy.

    \item \textbf{Reinforcement Learning for Optimizing Cluster Longevity:} Proposal of a reinforcement learning (RL)-based agent to minimize control overhead by learning the optimal timing for generating new clusters. This mechanism addresses a major limitation of existing protocols, which typically rely on fixed re-clustering intervals, leading to excessive control overhead. By dynamically determining the appropriate timing for re-clustering, the RL agent strikes a balance between reducing overhead and maintaining network performance.

    \item \textbf{Comprehensive Performance Evaluation:} Comparative analysis of LEACH-RLC against five baseline protocols—LEACH, LEACH-C, EE-LEACH, LEACH-D, and LEACH-CM—under various network configurations and metrics, such as energy efficiency, control overhead, and network lifetime. This analysis highlights the advantages of LEACH-RLC and identifies specific areas where existing protocols fall short.

    \item \textbf{Addressing Key Research Questions:} Systematic investigation and answers to three critical research questions, which are central to the design of energy-efficient clustering protocols:
          \begin{itemize}
              \item \textbf{RQ1:} Can we effectively reduce control overhead without compromising network performance?
              \item \textbf{RQ2:} How frequently should the controller initiate the generation of a new clustering solution?
              \item \textbf{RQ3:} When is the opportune moment for the controller to trigger a new clustering solution?
          \end{itemize}
          These questions are explored through a combination of the proposed MILP-based clustering approach, RL-based timing optimization, and a detailed comparative performance evaluation.
\end{enumerate}

The remainder of this paper is organized as follows.
Section~\ref{sec:related_work} presents the related work.
Section~\ref{sec:system_model} describes the system model.
Section~\ref{sec:proposed_solution_clustering} presents the proposed work for clustering.
Section~\ref{sec:proposed_solution_for_overhead_reduction} presents the proposed work for control overhead minimization.
Section~\ref{sec:results} presents the results.
Finally, Section~\ref{sec:conclusion} concludes the paper.

\section{Related Work}
\label{sec:related_work}

Efficient data collection and transmission with minimal energy consumption are primary goals in sensor networks, especially critical in battery-powered nodes deployed in remote \acrshort{iot} locations.
While direct transmission is simple, its energy inefficiency over long distances is a concern.
Clustering protocols address this by grouping nodes, reducing energy consumption.
Over the past two decades, extensive research has focused on designing and implementing clustering protocols for \acrshort{iot}.

One pioneering clustering protocol is \acrshort{leach}~\cite{heinzelmanEnergyefficientCommunicationProtocol2002}.
Nodes autonomously decide to be a \acrshort{ch} or cluster member based on a threshold and random number.
Expected numbers of \acrshortpl{ch} are determined by network size and desired \acrshort{ch} percentage $p$, with nodes expected to become \acrshortpl{ch} every $1/p$ round.
To address \acrshort{leach}'s limitations, \acrshort{leach-c} aims to find the optimal set of \acrshortpl{ch}, minimizing node energy consumption~\cite{heinzelmanApplicationspecificProtocolArchitecture2002a}.
A plethora of clustering protocols have stemmed from \acrshort{leach} and \acrshort{leach-c}.

Below, we present a review of recent clustering protocols in \acrshort{iot}, focusing on distributed and centralized approaches, and compare them with \acrshort{leach-rlc}.
Interested readers in a more comprehensive review of clustering protocols in \acrshort{iot} are referred to~\cite{daanouneComprehensiveSurveyLEACHbased2021,singhSurveySuccessorsLEACH2017,rawatClusteringProtocolsWireless2021}.

\subsection{Distributed Clustering Protocols}
\textit{Behera et al.} enhanced \acrshort{leach} by modifying the threshold function to improve energy efficiency~\cite{beheraResidualEnergybasedClusterhead2019a}.
Threshold calculation considers node initial energy, residual energy, and the optimal number of \acrshortpl{ch}.
Higher residual energy nodes are more likely to become \acrshortpl{ch}, preventing early network death.
\textit{Fathy et al.} proposed an adaptive data reduction technique for minimizing communication costs in IoT~\cite{fathyQualitybasedEnergyefficientData2019a}, using fine-grained sensor readings to reconstruct data and employing a dual prediction model.
\textit{Batta et al.} introduced an energy optimization clustering technique considering the \acrfull{soh} of sensor nodes' batteries~\cite{battaBatteryStateofhealthPredictionbased2022}, utilizing \acrshort{soh} to determine \acrshort{ch} sets less prone to battery degradation.
\textit{Chithaluru et al.} suggested an adaptive fuzzy-based, cluster-based routing protocol for IoT~\cite{chithaluruEnergyEfficientRoutingScheduling2021}, employing a fuzzy logic controller for volunteer node selection.
\textit{Behera et al.} improved the Stable Election Protocol~\cite{beheraISEPImprovedRouting2019}, incorporating unique threshold strategies and reducing control overhead.
\textit{Chen et al.} proposed a 2-hop clustering protocol, considering the distance and residual energy of nodes to select \acrshortpl{ch}~\cite{chenD2CRPNovelDistributed2022}.

\textit{Lee et al.} presented an energy-harvesting-aware clustering protocol for mobile \acrshort{wsn}~\cite{leeExtendedHierarchicalClustering2020,leeEnhancedHierarchicalClustering2017} using a two-tier fuzzy inference system.
\textit{Ahmad et al.} proposed LEACH-MEEC, optimizing energy consumption in mobile scenarios~\cite{ahmadNovelConnectivityBasedLEACHMEEC2018}.
\textit{Mohapatra et al.} suggested energy-efficient clustering protocols for mobile scenarios~\cite{mohapatraMobilityInducedMultihop2022}, extending LEACH to multi-path LEACH protocols.
\textit{Bharany et al.} proposed EE-LEACH, a modified LEACH protocol for energy efficiency~\cite{bharanyEnergyEfficientClusteringScheme2021a}, considering residual energy, drain rate, and network average measurements of both energy and draining rates.
\textit{Liu et al.} introduced LEACH-D, an adaptation of the classic \acrshort{leach} algorithm, which addresses the early exhaustion of certain \acrshortpl{ch} positioned at a significant distance from the \acrshort{bs}.
LEACH-D involves a second clustering stage where one or more CHs are selected as relay nodes for data fusion and transmission, significantly improving energy efficiency and network lifetime~\cite{liuLEACHDLowenergyLowdelay2024a}.
\textit{Tadros et al.} presented a modified LEACH-based clustering protocol using K-means for energy-efficient water quality monitoring, focusing on lightweight machine learning for extended network lifetime and accurate pollutant detection~\cite{tadrosUnsupervisedLearningBasedWSN2023}.
\textit{Qu et al.} introduced the Energy-Efficient Dynamic Clustering (EEDC), a distributed protocol for event monitoring in large-scale \acrshortpl{wsn}, employing Rough Fuzzy C-Means (RFCM) for overlapping cluster formation and a Genetic Algorithm (GA) for dynamic cluster head selection, achieving high energy efficiency and adaptability to event development~\cite{quEnergyEfficientDynamicClustering2021}.
\textit{Sun et al.} an adaptive clustering routing protocol for underwater sensor networks using multi-agent \acrshort{rl} to optimize global route selection. \acrshort{rl}  is employed to enable nodes to collaboratively select energy-efficient routes and adaptively decide cluster head roles based on environmental and routing data, improving network efficiency and lifetime~\cite{sunAdaptiveClusteringRouting2022}.
in~\cite{gururajCollaborativeEnergyEfficientRouting2023a} the authors proposed a a Collaborative Energy-Efficient Routing Protocol (CEEPR) for 5G/6G \acrshortpl{wsn}.
\acrshort{rl} is used to cluster network nodes and optimize cluster head selection based on residual energy, ensuring efficient data transmission.
Additionally, a multi-objective improved seagull algorithm (MOISA) enhances routing performance.
The authors in~\cite{aryaPerformanceAnalysisDeep2022a} proposed an energy-efficient routing protocol for \acrshort{iot}, combining \acrshort{rl}, Mantaray Foraging Optimization (MRFO), and deep belief networks (DBNs). \acrshort{rl} is used for clustering, where sensor nodes act as learning agents analyzing energy levels of neighbors to form clusters. \acrshortpl{ch} are selected using MRFO based on updated Q-values. DBNs are employed for efficient data transmission from CHs to the sink.

\subsection{Centralized Clustering Protocols}

\acrshort{leach-c} has been foundational for other centralized clustering protocols.
\textit{Zhang et al.} proposed a centralized clustering protocol for mobile WSN~\cite{zhangCentralizedEnergyEfficientClustering2019}, aimed at reducing node energy consumption while maximizing \acrfull{pdr}.
\textit{Chen et al.} proposed a randomly centralized clustering protocol to alleviate workload~\cite{chenRANCERandomlyCentralized2022}, enabling bidirectional heartbeat messages for event-driven, on-demand cluster creation.
\textit{Tebessia et al.} proposed an energy consumption model on a centralized controller to estimate node energy consumption~\cite{tebessiImprovementLEACHCProtocol2022}.
\textit{Gamal et al.} introduced a hybrid PSO and K-means clustering strategy~\cite{gamalEnhancingLifetimeWireless2022}, extending network lifetime and improving stability.
\textit{Jin et al.} presented a multi-hop routing protocol for \acrshort{wsn}~\cite{jinCentralizedMultiHopRouting2021}, implementing the minimum spanning forest algorithm for intra-cluster and inter-cluster routing tree construction.
\textit{Ma et al.} proposed an adaptive Voronoi diagram-based clustering protocol~\cite{maESCVADEnergySavingRouting2021} and a \acrshort{ch} selection algorithm based on the weighted sum of the distance to the \acrshort{ch} and the residual energy of the nodes.
\textit{Parmar et al.} introduced a modified LEACH-C (LEACH-CM) protocol that considers the distance between the selected \acrshort{ch} and a member node, and the distance between the member node and \acrshort{bs} to transmit data. LEACH-CM selects the number of \acrshortpl{bs} based on the alive nodes in the network, rather than the total number of nodes, showing improved network lifetime over LEACH-C~\cite{parmarImprovedModifiedLEACHC2016}.
\textit{Hassan et al.} proposed an improved energy-efficient clustering protocol (IEECP), a centralized clustering protocol that improves energy efficiency and prolongs the network lifetime by using a modified fuzzy C-means (M-FCM) algorithm for balanced-static clusters and a \acrshort{ch} selection-rotation algorithm to optimize energy consumption~\cite{hassanImprovedEnergyefficientClustering2020}.
\textit{Gurumoorthy et al.} proposed a secure and energy-aware routing protocol for \acrshort{wsn} using \acrfull{dcnn} for energy prediction and \acrfull{beassa} for optimal \acrshort{ch} selection, achieving high \acrshort{pdr} and improved network efficiency~\cite{gurumoorthyOptimalClusterHead2022}.
In~\cite{zhuECRKQMachineLearningBased2021a}, the authors proposed an energy-efficient clustering and routing protocol combining K-means and Q-learning. \acrshortpl{ch} are selected using K-means based on node residual energy and distance to the cluster centroid. Q-learning is used to associate nodes with \acrshortpl{ch}, optimizing the Q-value function to account for \acrshort{ch} residual energy and energy consumption during data transmission to the \acrshort{ch} and the base station.
The authors in~\cite{sathyamoorthyImprovedKMeansBased2021} proposed an optimal Q-learning-based clustering and load-balancing technique using an improved K-means algorithm. In the clustering phase, sensor nodes are assigned to clusters using Q-learning, and \acrshortpl{ch} are selected based on residual energy and distance to the sink. The node balancing phase evenly distributes sensors across partitions within clusters, with partition heads elected based on residual energy. \acrshortpl{ch} are dynamically re-elected when their energy falls below a threshold. The approach optimizes rewards by improving throughput, end-to-end delay, packet delivery ratio, and energy consumption.
In~\cite{al-jerewReinforcementLearningDelay2023a}, the authors proposed the Bounded Hop Count-Reinforcement Learning Algorithm (BHC-RLA) for optimizing \acrshort{ch} selection in \acrshortpl{wsn}. Using Q-learning, the algorithm selects \acrshortpl{ch} by maximizing a reward function that balances energy efficiency, data-gathering latency, and mobile \acrshort{bs} tour length. BHC-RLA constructs clusters with bounded hop counts, minimizes isolated nodes, and merges clusters to improve network performance.

\begin{table*}[ht!]
    \centering
    \caption{Comparison of recent clustering protocols in IoT}
    \label{tab:comparison}
    \def\arraystretch{1}%
    \begin{NiceTabular}[c]{ccccm{11.7cm}}[vlines]
        \CodeBefore
        \rowcolors{3}{gray!12}{}[respect-blocks]
        \Body
        \toprule
        \RowStyle[nb-rows=3,rowcolor=lightgray]{\bfseries}
        \Block{2-1}{Article}                                                                & \Block{2-1}{Year} & \Block{2-1}{AI/ML} & Control  & \Block{2-1}{Key Features}                                                                                                                                        \\
                                                                                            &                   &                    & Overhead &                                                                                                                                                                  \\
        \midrule
        \Block{1-5}{Distributed Clustering Protocols}                                                                                                                                                                                                                                                              \\
        \midrule
        \cite{ahmadNovelConnectivityBasedLEACHMEEC2018}                                     & 2018              & \xmark             & \xmark   & LEACH-MEEC for mobile scenarios                                                                                                                                  \\
        \cite{beheraResidualEnergybasedClusterhead2019a}                                    & 2019              & \xmark             & \xmark   & Modified LEACH threshold                                                                                                                                         \\
        \cite{fathyQualitybasedEnergyefficientData2019a}                                    & 2019              & \xmark             & \xmark   & Adaptive data reduction                                                                                                                                          \\
        \cite{beheraISEPImprovedRouting2019}                                                & 2019              & \xmark             & \cmark   & Improved Stable Election Protocol                                                                                                                                \\
        \cite{leeExtendedHierarchicalClustering2020, leeEnhancedHierarchicalClustering2017} & 2020              & \xmark             & \xmark   & Energy-harvesting-aware clustering                                                                                                                               \\
        \cite{chithaluruEnergyEfficientRoutingScheduling2021}                               & 2021              & \xmark             & \xmark   & Fuzzy-based, cluster-based routing                                                                                                                               \\
        \cite{quEnergyEfficientDynamicClustering2021}                                       & 2021              & \cmark             & \xmark   & RFCM for overlapping clusters, GA for dynamic CH selection, adaptability to event development, high energy efficiency.                                           \\
        \cite{bharanyEnergyEfficientClusteringScheme2021a}                                  & 2021              & \xmark             & \xmark   & EE-LEACH for energy efficiency (residual energy, drain rate, and network average measurements)                                                                   \\
        \cite{battaBatteryStateofhealthPredictionbased2022}                                 & 2022              & \xmark             & \xmark   & Energy optimization considering SOH                                                                                                                              \\
        \cite{mohapatraMobilityInducedMultihop2022}                                         & 2022              & \xmark             & \xmark   & Multi-path LEACH for mobility                                                                                                                                    \\
        \cite{chenD2CRPNovelDistributed2022}                                                & 2022              & \xmark             & \xmark   & 2-hop clustering protocol                                                                                                                                        \\
        \cite{sunAdaptiveClusteringRouting2022}                                             & 2022              & \cmark             & \xmark   & Adaptive clustering routing protocol using multi-agent \acrshort{rl} for global route selection and energy-efficient routing.                                    \\
        \cite{aryaPerformanceAnalysisDeep2022a}                                             & 2022              & \cmark             & \xmark   & Energy-efficient routing protocol combining \acrshort{rl}, MRFO, and DBNs for clustering and data transmission optimization.                                     \\
        \cite{gururajCollaborativeEnergyEfficientRouting2023a}                              & 2023              & \cmark             & \xmark   & Collaborative Energy-Efficient Routing Protocol for 5G/6G \acrshortpl{wsn} using \acrshort{rl} for cluster formation and optimal \acrshort{ch} selection.        \\
        \cite{tadrosUnsupervisedLearningBasedWSN2023}                                       & 2023              & \cmark             & \xmark   & Modified K-means clustering for energy efficiency and network lifetime                                                                                           \\
        \cite{liuLEACHDLowenergyLowdelay2024a}                                              & 2024              & \xmark             & \xmark   & A two-hierarchy clustering protocol for energy efficiency and network lifetime                                                                                   \\
        \midrule
        \RowStyle[nb-rows=1,rowcolor=lightgray]{\bfseries}
        \Block{1-5}{Centralized Clustering Protocols}                                                                                                                                                                                                                                                              \\
        \midrule
        \cite{parmarImprovedModifiedLEACHC2016}                                             & 2016              & \xmark             & \xmark   & Improved \acrshort{leach-c} with distance-based data transmission                                                                                                \\
        \cite{maESCVADEnergySavingRouting2021}                                              & 2019              & \xmark             & \xmark   & Reducing energy consumption in mobile WSN                                                                                                                        \\
        \cite{hassanImprovedEnergyefficientClustering2020}                                  & 2020              & \xmark             & \cmark   & Modified FCM for balanced clusters, centralized CH selection with back-off timer and rotation, energy balancing across CHs, suitable for long-lifetime networks. \\
        \cite{jinCentralizedMultiHopRouting2021}                                            & 2021              & \xmark             & \xmark   & Multi-hop routing for WSN                                                                                                                                        \\
        \cite{maESCVADEnergySavingRouting2021}                                              & 2021              & \xmark             & \xmark   & Adaptive Voronoi diagram-based clustering                                                                                                                        \\
        \cite{zhuECRKQMachineLearningBased2021a}                                            & 2021              & \cmark             & \xmark   & Energy-efficient clustering and routing protocol combining K-means and Q-learning for optimal \acrshort{ch} selection and node-\acrshort{ch} association.        \\
        \cite{sathyamoorthyImprovedKMeansBased2021}                                         & 2022              & \cmark             & \xmark   & Q-learning-based clustering and load-balancing technique using an improved K-means algorithm for optimal \acrshort{ch} selection and node balancing.             \\
        \cite{chenRANCERandomlyCentralized2022}                                             & 2022              & \xmark             & \cmark   & Randomly centralized clustering                                                                                                                                  \\
        \cite{tebessiImprovementLEACHCProtocol2022}                                         & 2022              & \xmark             & \xmark   & Energy consumption model estimation                                                                                                                              \\
        \cite{gamalEnhancingLifetimeWireless2022}                                           & 2022              & \xmark             & \xmark   & Hybrid PSO and K-means strategy                                                                                                                                  \\
        \cite{gurumoorthyOptimalClusterHead2022}                                            & 2022              & \cmark             & \cmark   & \acrfull{dcnn} for energy prediction and \acrfull{beassa} for \acrshort{ch} selection                                                                            \\
        \cite{al-jerewReinforcementLearningDelay2023a}                                      & 2023              & \cmark             & \xmark   & Bounded Hop Count-Reinforcement Learning Algorithm for \acrshort{ch} selection in \acrshortpl{wsn}                                                               \\
        \hline
        \RowStyle[rowcolor=lightgray]{\bfseries}
        \acrshort{leach-rlc}                                                                & 2025              & \cmark             & \cmark   & \acrshort{milp} for \acrshort{ch} selection and assignment, and \acrshort{rl} for control overhead minimization                                                  \\
        \bottomrule
    \end{NiceTabular}
\end{table*}

\subsection{Comparison with Related Work}

To provide a comprehensive overview of clustering protocols in \acrshort{iot}, Table \ref{tab:comparison} presents a comparative analysis of various approaches, highlighting key aspects such as the year of publication, incorporation of AI/ML techniques, handling of control overhead, and distinctive features of each protocol.

The significance of our work, \acrshort{leach-rlc}, is underscored by its unique combination of features that directly address the limitations observed in existing clustering protocols for \acrshort{iot} networks.
While prior protocols, such as those presented in the literature, have struggled to incorporate both optimal \acrshort{ch} selection and efficient control overhead management simultaneously, \acrshort{leach-rlc} stands out as a pioneering solution.

Interestingly, while many protocols have improved energy efficiency, and network lifetime, they often lack adaptability to dynamic network conditions and optimal timing for generating new clusters.
The majority of existing protocols (\cite{beheraResidualEnergybasedClusterhead2019a}, \cite{fathyQualitybasedEnergyefficientData2019a}, \cite{battaBatteryStateofhealthPredictionbased2022}, \cite{chithaluruEnergyEfficientRoutingScheduling2021}, \cite{leeExtendedHierarchicalClustering2020, leeEnhancedHierarchicalClustering2017}, \cite{ahmadNovelConnectivityBasedLEACHMEEC2018}, \cite{mohapatraMobilityInducedMultihop2022}, \cite{zhangCentralizedEnergyEfficientClustering2019}, \cite{tebessiImprovementLEACHCProtocol2022}, \cite{jinCentralizedMultiHopRouting2021}, \cite{maESCVADEnergySavingRouting2021}, \cite{chenD2CRPNovelDistributed2022}) have primarily focused on specific aspects, such as modified thresholds, energy optimization, or clustering strategies, but have fallen short in simultaneously optimizing both energy efficiency and control overhead.
Additionally, while some protocols have incorporated AI/ML techniques, such as \acrshort{rl} (\cite{tadrosUnsupervisedLearningBasedWSN2023, sunAdaptiveClusteringRouting2022, aryaPerformanceAnalysisDeep2022a, gururajCollaborativeEnergyEfficientRouting2023a,zhuECRKQMachineLearningBased2021a,sathyamoorthyImprovedKMeansBased2021,al-jerewReinforcementLearningDelay2023a}), they have not fully addressed the challenges of optimal timing for generating new clusters and minimizing control overhead.

\acrshort{leach-rlc} introduces a centralized clustering protocol that leverages \acrshort{milp} for optimal \acrshort{ch} selection and node-cluster assignments, integrating an \acrshort{rl} agent designed to minimize control overhead by learning the optimal timing for sending control messages.

\subsubsection{The Need for AI/ML Techniques}
Traditional protocols often rely on static or heuristic methods for clustering and control message scheduling, which do not adapt well to dynamic network conditions. AI, particularly \acrshort{rl}, offers the ability to learn from the network environment and make real-time decisions, providing a more responsive and efficient clustering mechanism.
Additionally, solving the \acrshort{milp} problem is computationally expensive and unsuitable for training in real-time systems.
Trained neural networks can expedite this process, making it feasible for real-time use.

\subsubsection{The Role of AI/ML}
The \acrshort{rl} agent in \acrshort{leach-rlc} continuously monitors the network state and learns to optimize the timing and frequency of control messages, reducing unnecessary exchanges and conserving energy. By utilizing neural networks to solve the \acrshort{milp} problem, \acrshort{leach-rlc} can quickly be trained to determine the optimal \acrshort{ch} set and node-cluster assignments, reducing computational complexity and enabling real-time operation.

\subsubsection{The Significance of AI}
The integration of AI in \acrshort{leach-rlc} addresses the critical gap in existing protocols, which often fail to balance energy efficiency and control overhead effectively. The \acrshort{rl} agent's ability to adaptively manage control overhead significantly reduces the energy consumption associated with control message exchanges, a major limitation of traditional clustering protocols. The use of a neural network for predicting cluster configurations ensures rapid and efficient decision-making, enhancing the protocol's capability to operate in real-time systems.

This innovative application of AI not only improves the performance and efficiency of \acrshort{leach-rlc} but also sets a precedent for future \acrshort{iot} clustering protocols to leverage AI for dynamic and optimized network management. The incorporation of AI in \acrshort{leach-rlc} is not just a superficial addition but a fundamental innovation that enhances the protocol's efficiency, adaptability, and overall performance.

\section{System Model}
\label{sec:system_model}

In our \acrshort{iot} network, $N$ nodes are randomly deployed within a $L \times L$ area.
Each node is equipped with an initial energy supply of $E_{0}$ and is identified by a unique identifier, $i$, along with coordinates $(x_i, y_i)$.

The communication graph $G = (N, \mathcal{E})$ represents the network, where $N$ is the set of nodes and $\mathcal{E}$ is the set of edges.
Nodes collect data from the environment and transmit it to their respective \acrshortpl{ch}.
\acrshortpl{ch} aggregate the data from their cluster members and send the aggregated data to the \acrshort{bs}, which is directly connected to the centralized controller.
The controller is typically located at the \acrshort{bs} and is responsible for managing the network.
It is deployed in a fixed location, such as a dedicated server room or a protected cabinet near the \acrshort{bs}, ensuring it is well-maintained and accessible for any required updates or maintenance activities.
For each $n \in N$, the status containing the energy level $E_{i}$ is updated at the end of each round and shared with the controller.
Also, initial locations and energy levels are shared with the controller at the beginning of the network operation.
The controller maintains a global view of the network by continuously updating its database with the status information received from the nodes.
This includes energy levels, node positions, and any other relevant operational data.
By having a comprehensive and up-to-date network view, the controller can make informed decisions about network management and optimization.

\begin{figure}[!t]
    \centering
    \includegraphics[width=1\columnwidth]{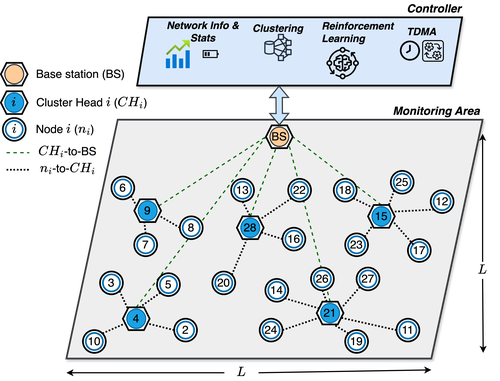}
    \caption{System model of the \acrshort{iot} network.}
    \label{fig:system_model}
\end{figure}

The controller, equipped with a global network view, makes decisions on the optimal timing for creating new clusters, selecting \acrshortpl{ch}, and assigning nodes to clusters.
Additionally, the controller manages the distribution of \acrshort{tdma} schedules to nodes.
Fig.~\ref{fig:system_model} illustrates the system model, highlighting the interactions between the nodes, \acrshortpl{ch}, \acrshort{bs}, the controller, and the functions performed by the controller.

The key assumptions in our model are as follows:
\begin{enumerate}
    \item Nodes are static: Nodes remain in fixed positions throughout the network operation; mobility is not considered in this study.
    \item Each node is aware of its own location: The exact locations of the nodes are assumed to be known by the controller, which aids in solving the \acrshort{milp}.
          This assumption could be relaxed in practical deployments by using location estimation methods.
    \item Nodes can adjust their transmission power settings.
    \item The \acrshort{bs} and controller remain stationary and are not resource-constrained after deployment.
\end{enumerate}

We consider these assumptions reasonable for typical \acrshort{iot} applications.
For example, in a smart city application, the \acrshort{bs} and controller could be deployed in a fixed location, such as a building or a pole.
Nodes, situated in the streets, may be powered by solar energy and equipped with \acrfull{gps} receivers to determine their locations.
Table~\ref{tab:notation} summarizes the notation used throughout the paper.

\begin{table}
    \centering
    \caption{Notation}
    \label{tab:notation}
    \def\arraystretch{1}%
    \begin{NiceTabular}[c]{lm{7cm}}[vlines]
        \CodeBefore
        \rowcolor{lightgray}{1}
        \rowcolors{2}{gray!12}{}[respect-blocks]
        \Body
        \toprule
        \RowStyle[]{\bfseries}
        Symbol           & Description                                                                                                                                  \\
        \midrule
        $N$              & \(N=\{n_1, n_2, \ldots, n_{|N|}\}\) is the set of nodes in the network                                                                       \\
        $L$              & Length of the area                                                                                                                           \\
        $E_{0}$          & Initial energy supply of nodes                                                                                                               \\
        $E_{i}$          & Energy level of node $i$                                                                                                                     \\
        $E^d_i$          & \(E^d_i = E^{r-1}_i - E^{r}_i\) is the energy dissipated by node $i$, and $r$ and $r-1$ denote the current and previous rounds, respectively \\
        $\bar{E_{d}}$    & \(\bar{E_{d}} = \frac{1}{|N|}\sum_{i \in N} E^d_i\) is the average energy dissipated by the nodes in the network                             \\
        $E_{net}$        & \(E_{net} = \sum_{i \in N} E_{i}\) is the total energy of the network at round $r$                                                           \\
        $\bar{E}$        & \( \bar{E} = \frac{E_{net}}{|N|}\) is the average energy level of the network                                                                \\
        $\mathcal{D} $   & \( \mathcal{D} = \{n \in N | E_n > 0\},~\mathcal{D} \subseteq N \) is the set of nodes currently active or operational                       \\
        $H$              & \(H=\{n \in N | E_n \geq \bar{E}\},~H \subseteq N \) is the set of potential cluster heads                                                   \\
        $C_{i}$          & Cluster of node $i$                                                                                                                          \\
        $CH$             & \(CH = \{CH_1, CH_2, \ldots, CH_{|N|}\},~CH \subset N \) is the set of cluster heads                                                         \\
        $CH_{\tau}$      & Number of rounds elapsed since the last cluster head selection                                                                               \\
        $CA$             & Current cluster assignment                                                                                                                   \\
        $E_{elec}$       & Energy consumed by the transmitter/receiver circuitry per bit                                                                                \\
        $E_{fs}$         & Energy parameter for the free space model                                                                                                    \\
        $E_{amp}$        & Energy parameter for the multi-path fading model                                                                                             \\
        $d_{0}$          & Threshold distance determining the transition between the free space and multi-path fading models                                            \\
        $d_{ij}$         & Distance between nodes $i$ and $j$                                                                                                           \\
        $B_{i}$          & Packet size of node $i$                                                                                                                      \\
        $B_{c}$          & Control packet size                                                                                                                          \\
        $k$              & Percentage of nodes that are allowed to become \acrshortpl{ch}                                                                               \\
        $E^{(i,j)}_{tx}$ & Energy consumed by node $i$ when transmitting a packet of size $B_{i}$ to node $j$                                                           \\
        $E_{rx}^i$       & Energy consumed by node $i$ when receiving a packet of size $B_{i}$                                                                          \\
        $E_{rx}^{ch_i}$  & Energy consumed by \acrshort{ch} $i$ when receiving data from its cluster members                                                            \\
        $E_{tx}^{ch_i}$  & Energy consumed by \acrshort{ch} $i$ when transmitting data to the \acrshort{bs}                                                             \\
        $E_{DA}$         & Energy consumed by data aggregation                                                                                                          \\
        $E_{rx}^{c_i}$   & Energy consumed by node $i$ when receiving a control packet of size $B_{c}$                                                                  \\
        $\alpha$         & \(\alpha \in \mathbb{R}^+_0\) is a weighting factor for the energy consumed by nodes when transmitting to \acrshortpl{ch}.                   \\
        $\beta$          & \(\beta \in \mathbb{R}^+_0\) is a weighting factor for the energy consumed by \acrshortpl{ch} when transmitting to the sink.                 \\
        $\gamma$         & \(\gamma \in \mathbb{R}^+_0\) is a weighting factor for the energy consumed by \acrshortpl{ch} when receiving from their cluster members.    \\
        \bottomrule
    \end{NiceTabular}
\end{table}

\subsection{Energy Consumption Model}
\label{subsec:energy_consumption_model}

To estimate the energy consumption of nodes, we adopt the widely recognized energy consumption model for \acrshort{wsn} nodes proposed in~\cite{heinzelmanEnergyefficientCommunicationProtocol2002}.
This model integrates both the free space and multi-path fading models, selecting the appropriate model based on the distance between the transmitter and the receiver.

The threshold distance ($d_{0}$) determining the transition between the free space and multi-path fading models is calculated as follows:
\begin{equation}
    d_{0} = \sqrt{\frac{E_{fs}}{E_{amp}}},
\end{equation}
where $E_{fs}$ represents the energy parameter for the free space model, and $E_{amp}$ is the energy parameter for the multi-path fading model.

The energy consumed by a node $i$ when transmitting a packet of size $B_{i}$ to a node $j$ is given by:
\begin{equation}
    E^{(i,j)}_{tx} = \left\{
    \begin{array}{ll}
        E_{elec} \times B_{i} + E_{fs} \times B_{i} \times d_{ij}^{2}  & \mbox{if } d_{ij} \leq d_{0} \\
        E_{elec} \times B_{i} + E_{amp} \times B_{i} \times d_{ij}^{4} & \mbox{otherwise}
    \end{array}
    \right.
    ,
\end{equation}
where $E_{elec}$ is the energy consumed by the transmitter or receiver circuitry per bit, and $d_{ij}$ is the distance between nodes $i$ and $j$.

The energy consumed when receiving a packet of size $B_{i}$ is calculated as:
\begin{equation}
    E_{rx}^i = E_{elec} \times B_{i}.
\end{equation}

\acrshortpl{ch} incur additional energy consumption due to the overhead of receiving and transmitting data within their clusters. Specifically, the energy consumed by a \acrshort{ch} $i$ when receiving data from its cluster members is given by:
\begin{equation}
    E_{rx}^{ch_i} = E_{elec} \times \sum_{j \in C_{i}} B_{j},
\end{equation}
where $C_{i}$ denotes the set of nodes in the cluster of \acrshort{ch} $i$.

Moreover, \acrshortpl{ch} consume energy when transmitting data to the \acrshort{bs}.
This energy consumption is expressed as:
\begin{equation}
    E_{tx}^{ch_i} = (E_{elec} + E_{DA}) \times B_{i} + E_{tx},
\end{equation}
where $E_{DA}$ represents the energy consumed by data aggregation.
The network incurs additional energy consumption due to the overhead of transmitting control packets, such as \acrshort{ch} advertisements and \acrshort{ch} selection messages.
The energy consumed by a node $i$ when receiving a control packet of size $B_{c}$ is given by:
\begin{equation}
    E_{rx}^{c_i} = E_{elec} \times B_{c}.
\end{equation}

\section{Problem Formulation for Generating Clusters}
\label{sec:proposed_solution_clustering}

In this section, we present the problem formulation for generating clusters in \acrshort{leach-rlc}.
To address the challenges of \acrshort{ch} selection and node assignment, we formulate a \acrshort{milp} problem.
We introduce the decision variables to formulate the \acrshort{milp} problem.

\begin{table}[ht!]
    \centering
    \caption{Decision variables for the \acrshort{milp} problem.}
    \def\arraystretch{1}%
    \begin{NiceTabular}[c]{lm{5.5cm}}[hvlines]
        \CodeBefore
        \rowcolor{lightgray}{1}
        \rowcolors{2}{gray!12}{}[respect-blocks]
        \Body
        \RowStyle[]{\bfseries}
        Variables                            & Description                                                                       \\
        \(x_j\), \(j \in H\)                 & Binary variable indicating whether node \(j\) is selected as a \acrshort{ch}.     \\
        \(y_{ij}\), \(i \in N\),~\(j \in H\) & Binary variable indicating whether node \(i\) is assigned to \acrshort{ch} \(j\). \\
    \end{NiceTabular}
\end{table}

The resulting \acrshort{milp} problem is formulated as follows:

\begin{subequations}
    \label{eq:optimization_problem}
    \begin{align}
         & \min \alpha \sum_{i \in N} \sum_{j \in H} E_{tx}^{(i,j)} y_{ij} + \beta \sum_{j \in H} E_{tx}^{ch_j} x_j+ \gamma \sum_{j \in H}\sum_{i \in N}E^i_{rx}y_{ij} \label{eq:objective_function} \\
         & \sum_{j \in H} y_{ij} = 1, \text{ for each } i \in \text{N},  \label{eq:constraint_one_cluster_head}                                                                                      \\
         & \sum_{j \in H} x_j = k, \label{eq:constraint_numer_clusters}                                                                                                                              \\
         & y_{ij} \leq x_j, \text{ for each } i \in \text{N}, j \in H. \label{eq:constraint_consistency}
    \end{align}
\end{subequations}

The objective function~(\ref{eq:objective_function}) is designed to minimize the energy expended by non-\acrshortpl{ch} while transmitting to their respective \acrshortpl{ch}, the energy consumed by \acrshortpl{ch} during transmission to the sink, and the energy consumed by \acrshortpl{ch} during reception from their cluster members.
The weighting factors \(\alpha\), \(\beta\), and \(\gamma\) provide a mechanism for the \acrshort{milp} problem to prioritize different aspects of energy consumption.

The parameter \(\alpha\) scales the energy spent by non-\acrshortpl{ch} during transmission to \acrshortpl{ch}, while \(\beta\) adjusts the energy used by \acrshortpl{ch} when transmitting to the sink.
Meanwhile, \(\gamma\) influences the energy consumed by \acrshortpl{ch} during the reception from their cluster members.
These weighting factors enable fine-tuning of the optimization process according to specific energy consumption priorities.

The constraints (\ref{eq:constraint_one_cluster_head}), (\ref{eq:constraint_numer_clusters}), and (\ref{eq:constraint_consistency}) contribute to the coherence and effectiveness of the clustering strategy.
The first ensures that each node is assigned to exactly one \acrshort{ch}, the second controls the selection of the desired number of \acrshortpl{ch}, and the third maintains consistency between \(x_j\) and \(y_{ij}\), ensuring that if node \(j\) is not selected as a \acrshort{ch}, then node \(i\) cannot be assigned to \acrshort{ch} \(j\).

The \acrshort{milp} problem not only minimizes overall energy consumption but also ensures a balanced workload distribution among the \acrshortpl{ch}.
This comprehensive approach results in an energy-efficient and effective clustering strategy for IoT networks.

The majority of parameters in the \acrshort{milp} problem are static and derivable from the network model, including \(E_{\text{tx}}^{(i,j)}\) and \(E_{\text{rx}}\).
Dynamic parameters, such as  \(H=\{n \in N | E_n \geq \bar{E}\},~H \subseteq N \), representing the set of potential \acrshortpl{ch}, are updated during each round of the clustering process.
The only unknown parameters are the weighting factors \(\alpha\), \(\beta\), and \(\gamma\), whose values significantly influence the \acrshort{milp} problem's performance.
Next, we explore an approach to determine these crucial parameter values.

\begin{figure*}[!ht]
    \centering
    \subfloat[\label{fig:weigths_alpha_beta}]{\includegraphics[width=0.32\textwidth]{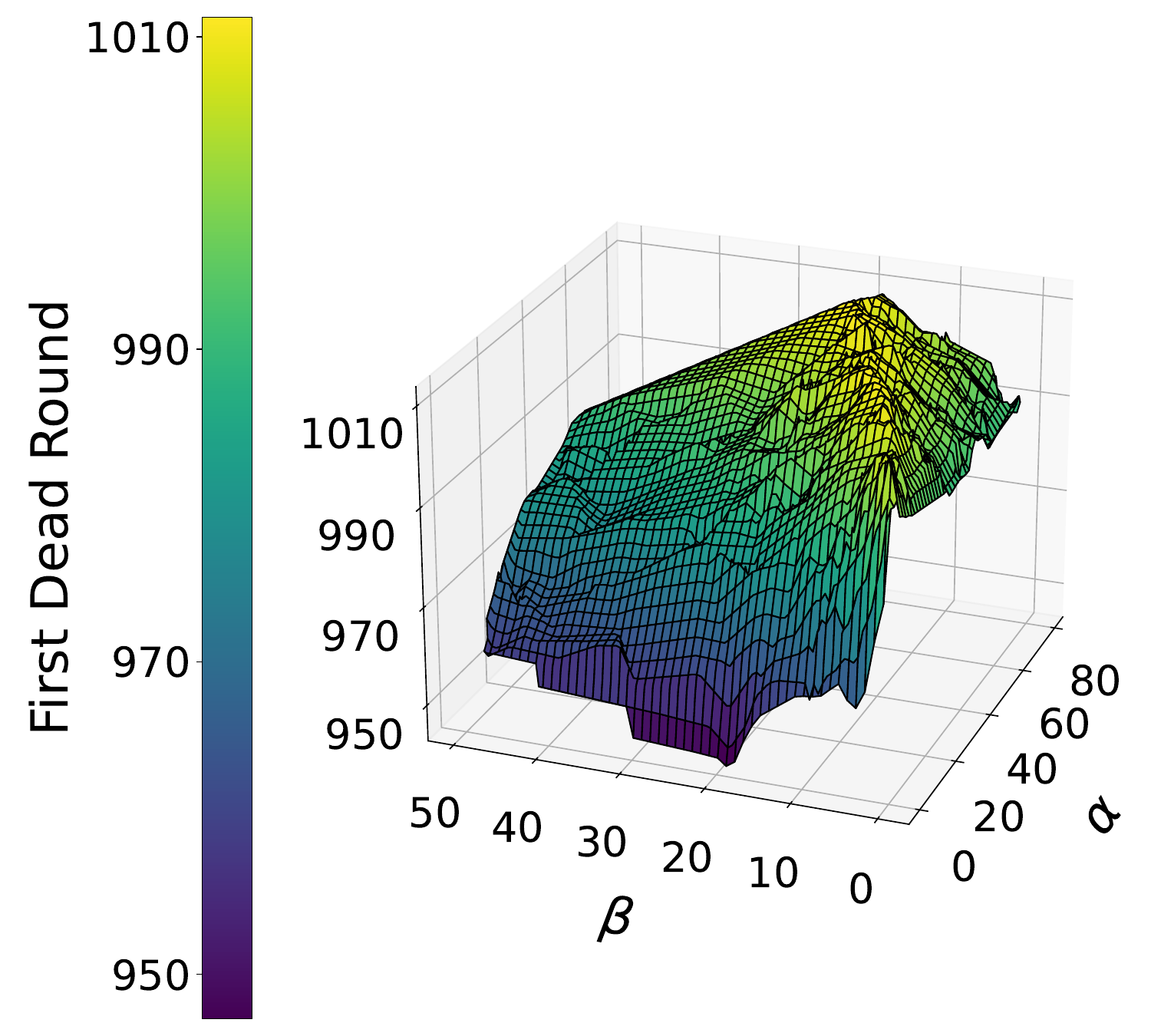}}
    \subfloat[\label{fig:weigths_alpha_gamma}]{\includegraphics[width=0.32\textwidth]{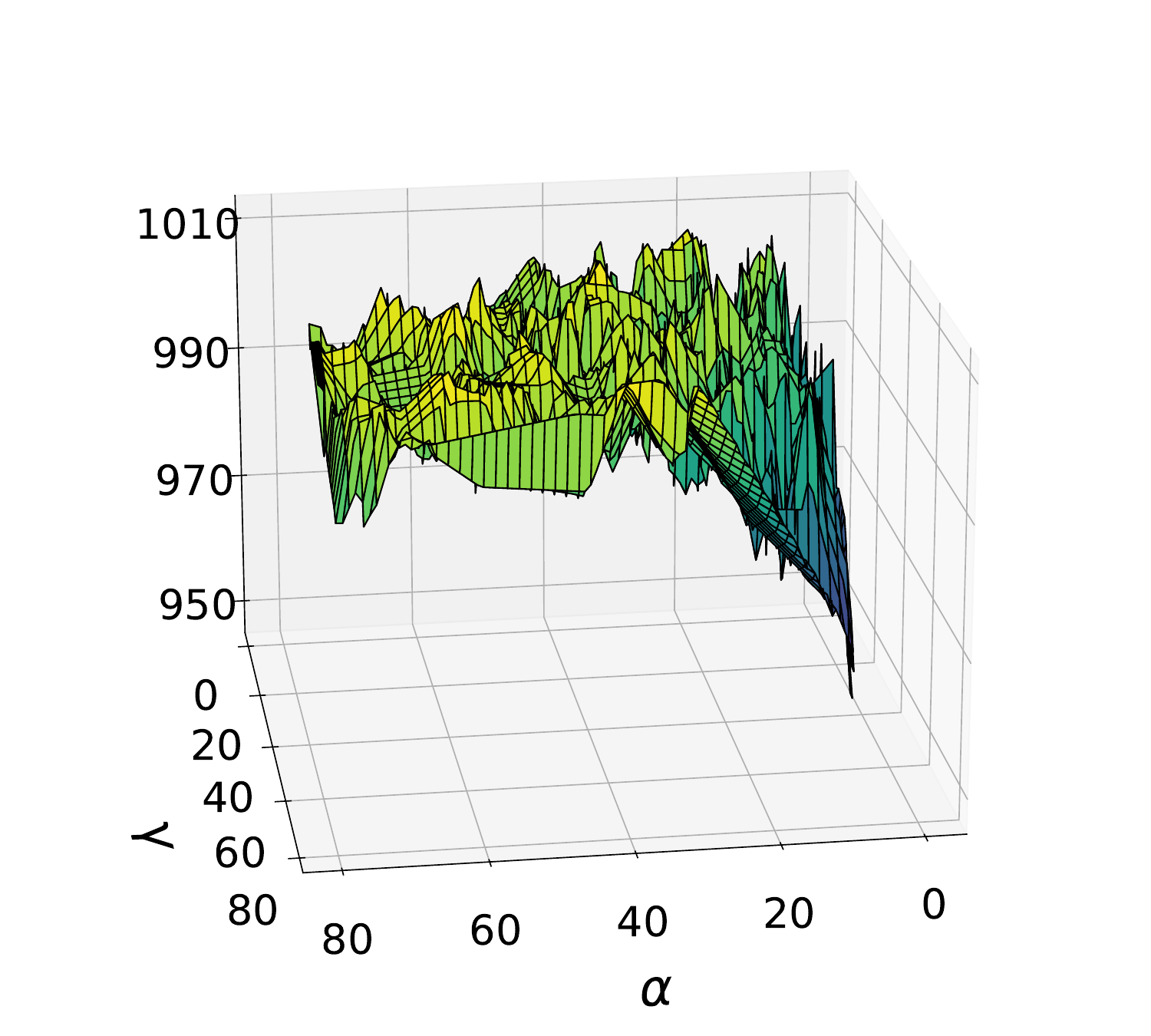}}
    \subfloat[\label{fig:weigths_beta_gamma}]{\includegraphics[width=0.32\textwidth]{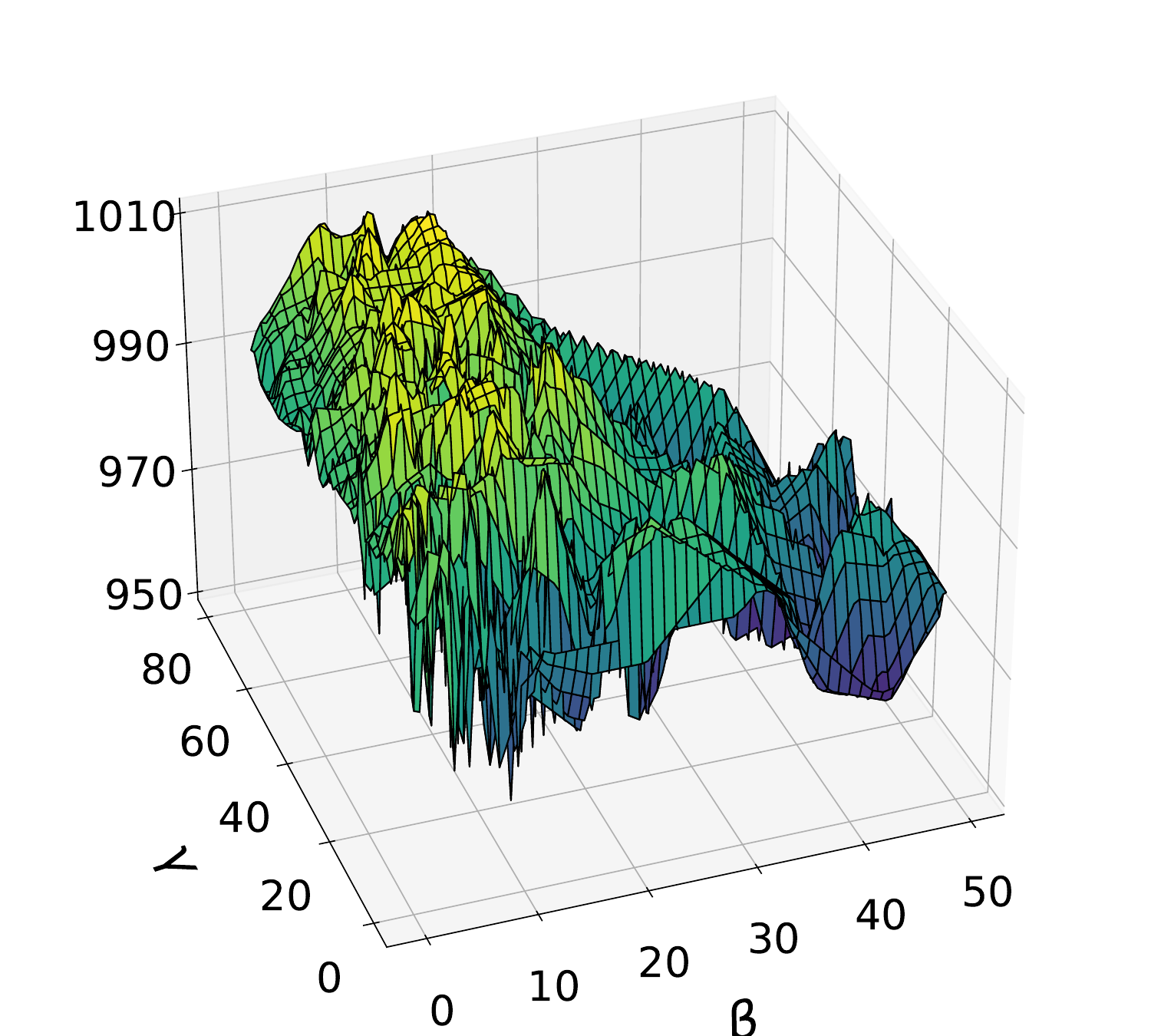}}
    \caption{Heatmap of the \acrshort{fnd} metric for different values of \(\alpha\), \(\beta\), and \(\gamma\).}
    \label{fig:heatmap}
\end{figure*}

\subsection{Weighting Factors Adjustment Strategies}

The optimal values for the weighting factors \(\alpha\), \(\beta\), and \(\gamma\) are highly dependent on the network's scale and node density.
In smaller networks with low node densities, the energy consumed by individual transmissions (\(\alpha\)) may dominate, necessitating a higher weight to reduce intra-cluster energy usage.
Conversely, in larger or denser networks, where the number of clusters and inter-cluster transmissions increase, \(\beta\) and  \(\gamma \) should be scaled up to account for the higher energy demands of \acrshortpl{ch}.
Additionally, in dense networks, careful adjustment of \(\gamma\) ensures that the energy cost of receiving data by \acrshortpl{ch} does not lead to rapid depletion of their resources.
A practical approach to determining these weights involves sensitivity analysis across different network topologies, where the performance metrics like \acrfull{fnd} and network lifetime are evaluated for various parameter combinations.
By identifying trends across scales and densities, the weighting factors can be adjusted to maintain energy efficiency and workload balance.

\subsection{Parameter Selection}
\label{subsec:parameter_selection}

The weighting factors \(\alpha\), \(\beta\), and \(\gamma\) are crucial to the performance of the \acrshort{milp} problem.
These parameters influence the optimization process and determine the clustering strategy's effectiveness.
The values of these parameters are not known \emph{a priori} and must be determined before the clustering process begins.
Here, we present an approach to determine the values of these parameters.

To determine the values for the weighting factors, we systematically evaluate the performance of the \acrshort{milp} problem for different values of \(\alpha\), \(\beta\), and \(\gamma\).
We utilized the \acrshort{milp} problem to generate clustering solutions for a total of 600 combinations of (\(\alpha\), \(\beta\), and \(\gamma\)), with each parameter ranging from 0 to 100.
The evaluation of clustering solutions was based on the \acrshort{fnd} metric, representing the number of rounds until the first node exhausts its energy.
Fig.~\ref{fig:heatmap} presents a heatmap illustrating the \acrshort{fnd} metric for diverse values of \(\alpha\), \(\beta\), and \(\gamma\).

\subsubsection{Heatmap of the FND metric}

The heatmap distinctly reveals the sensitivity of the \acrshort{fnd} metric to variations in \(\alpha\), \(\beta\), and \(\gamma\).
Noteworthy observations include the following:
\begin{itemize}
    \item  Fig.~\ref{fig:weigths_alpha_beta} and Fig.~\ref{fig:weigths_beta_gamma}, the \acrshort{fnd} metric attains its peak when  \(\beta<30\).
    \item  Fig.~\ref{fig:weigths_alpha_beta} and Fig.~\ref{fig:weigths_alpha_gamma} showcase that the \acrshort{fnd} metric is highest when \(\alpha > 20\).
    \item Fig.~\ref{fig:weigths_alpha_gamma} and Fig.~\ref{fig:weigths_beta_gamma} highlight that the \acrshort{fnd} metric achieves its maximum when \(\gamma > 30\).
\end{itemize}

\subsubsection{Optimal Parameter Values}

These observations underscore the sensitivity of the \acrshort{fnd} metric to the specific values of \(\alpha\), \(\beta\), and \(\gamma\).
The overall best performance is attained when employing the following parameter values: \(\alpha = 54.83\), \(\beta = 14.54\), and \(\gamma = 35.31\). Importantly, these values are tailored to the network scenario discussed in Section~\ref{sec:results}.

These findings provide crucial insights for selecting optimal weighting factor values, ensuring enhanced energy efficiency and network longevity in the context of the proposed \acrshort{milp}-based clustering strategy.

\section{Optimizing Clustering Timing and Frequency with Reinforcement Learning to Reduce Overhead}
\label{sec:proposed_solution_for_overhead_reduction}

In this section, we introduce our innovative solution crafted to mitigate control overhead in the network, paving the way for improved network performance.
We aim to address the following research questions:
\begin{itemize}
    \item \textbf{RQ1:} Can we effectively reduce control overhead without compromising network performance?
    \item \textbf{RQ2:} How frequently should the controller initiate the generation of a new clustering solution?
    \item \textbf{RQ3:} When is the opportune moment for the controller to trigger a new clustering solution?
\end{itemize}

\begin{figure}[!t]
    \centering
    \includegraphics[width=1\columnwidth]{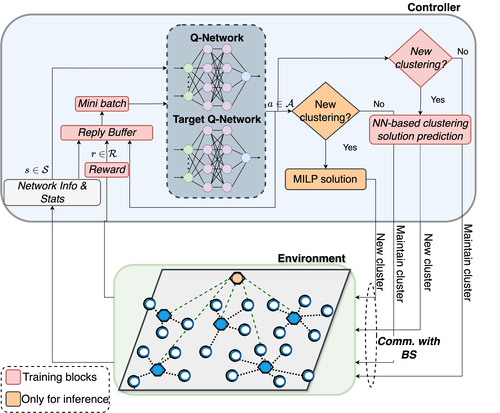}
    \caption{Proposed solution overview integrating the \acrshort{rl} agent and the MILP-based clustering solution.}
    \label{fig:proposed_solution_overview}
\end{figure}

The proposed \acrshort{rl} approach determines the optimal timing for re-clustering by learning from stochastic network dynamics, while the MILP formulation provides the best clustering configuration for a given snapshot of the network state, achieving a complementary balance between adaptability and efficiency.
Fig.~\ref{fig:proposed_solution_overview} illustrates the proposed solution, highlighting the integration of the \acrshort{rl} agent and the MILP-based clustering solution.

\subsection{Problem Context and Proposed Solution Overview}
\label{subsec:proposed_solution_overview}

In our quest to diminish control overhead, we propose the integration of a \acrshort{rl} agent.
This agent is designed to learn and discern the optimal timing for generating a new clustering solution, thereby streamlining the control process.
Through the use of \acrshort{rl}, our solution seeks to strike a harmonious balance between minimizing control overhead and enhancing overall network efficiency.
Readers interested in a comprehensive overview of \acrshort{rl} and its networking applications are referred to~\cite{szepesvariAlgorithmsReinforcementLearning2010,luongApplicationsDeepReinforcement2019,jurado-lassoSurveyMachineLearning2022,jurado-lassoHRLTSCHHierarchicalReinforcement2024a}.

We cast the challenge of minimizing control overhead as a \acrfull{mdp}.
The \acrshort{mdp} is characterized by a tuple $\langle \mathcal{S}, \mathcal{A}, \mathcal{R} \rangle$, where $\mathcal{S}$ denotes the set of states, $\mathcal{A}$ signifies the set of actions, and $\mathcal{R}$ encapsulates the reward function~\cite{jurado-lassoELISEReinforcementLearning2024a}.

The state $s_t$ at time $t$ is comprehensively described by the tuple $\langle E_{net_t},E_{n_t}~\forall n \in N, CH_t, CH_{\tau_t}, CA_t, a_{t-1} \rangle$.
Here, $t$ also denotes the current round.
In this representation, \(E_{net_t}=\sum_{n \in N} E_n\) is the network's residual energy at time $t$, $E_{n_t}~\forall n \in N$ captures the residual energy of individual nodes at time $t$, $CH_t$ denotes the set of cluster heads, $CH_{\tau_t}$ indicates the number of rounds elapsed since the last cluster head selection, $CA_t$ depicts the current cluster assignment, and $a_{t-1} \in \mathcal{A}$ denotes the action taken at time $t-1$.

Now, delving into the action space $\mathcal{A}$, the agent has the flexibility to choose from the following discrete actions: \textit{generate a new clustering solution} and \textit{maintain the current clustering solution}.
This action space is denoted as $\mathcal{A} = \{a_1, a_2\}$, where $a_1$ represents the discrete action of generating a new clustering solution, and $a_2$ corresponds to the discrete action of maintaining the current clustering solution.

The immediate reward function $\mathcal{R}(s, a)$ for the agent in the proposed \acrshort{mdp} is designed to encourage behaviors that enhance the network's performance.
At each time step, the agent is rewarded with 1 point if no nodes experience depletion in their energy, promoting the longevity of the network without node losses.
Additionally, the agent receives a small reward of 0.1 when it chooses to generate a new clustering solution ($a_1$).
This value was selected as a minor incentive to encourage exploration without overshadowing the primary objective of maximizing network lifetime and energy efficiency.
By keeping the value significantly smaller than the main rewards, the agent is driven to test alternative clustering configurations while maintaining a focus on the primary task.
Experimental results confirmed that this balance improved adaptability to dynamic network conditions without detracting from overall task performance.
This encourages the agent to explore new cluster configurations, facilitating adaptability to changing network conditions.
Lastly, a higher reward of 2 is granted when the episode concludes, marking the end of a round.
The episode concludes when the first node experiences energy depletion, reflecting the critical nature of node failures.
This substantial reward is intended to signify the conclusion of the learning episode, capturing the significance of sustaining the network without premature node losses.
Mathematically, the reward function is defined as follows:
\begin{equation}
    \label{eq:reward_function}
    \mathcal{R}(s_t, a_t) = \begin{cases}
        1   & \text{if } a_t = a_2 \text{ and } E_{n_t} > 0 \text{ for all } n \in N \\
        1.1 & \text{if } a_t = a_1 \text{ and } E_{n_t} > 0 \text{ for all } n \in N \\
        2   & \text{if } E_{n_t} \leq 0 \text{ for some } n \in N
    \end{cases}
    .
\end{equation}

The proposed solution leverages the \acrshort{milp} described in Section~\ref{sec:proposed_solution_clustering} to generate a clustering solution.
However, the computational complexity of the \acrshort{milp}, even though invoked only when the agent opts to generate a new clustering solution, poses a potential bottleneck due to high training epochs.
On a 2.3 GHz 8-Core Intel Core i9 processor with 16 GB of \acrshort{ram}, the \acrshort{milp} requires approximately $0.91 \pm 0.4$ seconds to generate a new clustering solution.
In contrast, the \acrshort{rl} agent, introduced later, achieves the same task in just $0.0069 \pm 0.0004$ seconds.
Recognizing the computational disparity, we address this challenge through a novel approach outlined in the subsequent section.

\begin{figure}[t]
    \centering
    \includegraphics[width=0.95\columnwidth]{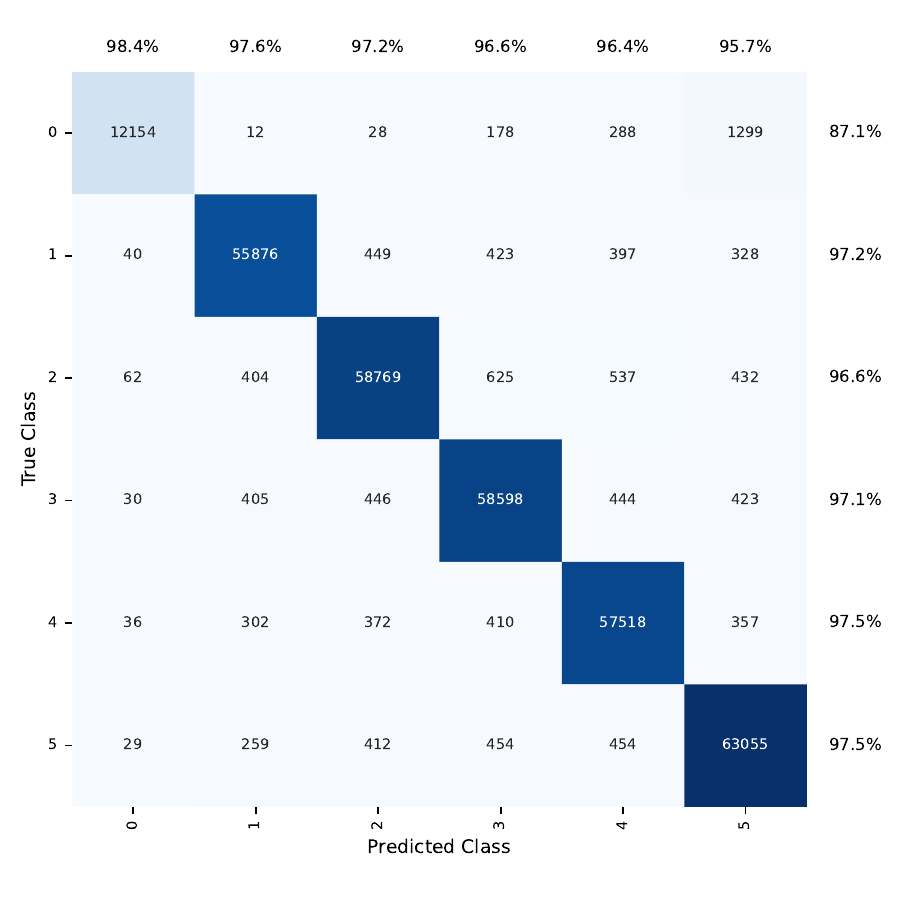}
    \caption{Confusion matrix for the neural network-based cluster member assignment.}
    \label{fig:ca_confusion_matrix}
\end{figure}

\subsection{Neural network-based clustering solution prediction}
\label{subsec:neural_network_based_clustering_solution_prediction}

To alleviate the computational complexity of the \acrshort{milp}, we advocate for the adoption of a neural network to predict the clustering solution.
This innovative approach aims to streamline the computational demands associated with clustering while maintaining accuracy.
The neural network is trained using the clustering solutions generated by the \acrshort{milp} as the ground truth.
However, we adopt a nuanced strategy by breaking down the clustering solution into two distinct components: the set of \acrshortpl{ch} and the set of cluster members.
The neural network is then trained to these two components separately.
This architecture enhances prediction accuracy, especially considering the key role of the \acrshortpl{ch}' feature in determining the cluster members.

\begin{algorithm}[!htb]
    \footnotesize %
    \caption{\acrshort{leach-rlc} training and evaluation algorithm}
    \label{alg:leach-rlc}
    \SetAlgoLined
    \SetKwComment{Comment}{\textcolor{blue}{/* }}{\textcolor{blue}{ */}}
    \SetKwInOut{Input}{Input}
    \SetKwInOut{Output}{Output}

    \Input{Network parameters, \(\alpha, \beta, \gamma, k, E_{net}, E_n~\forall n \in N, E^{ch_j}_{tx}, E^{(i,j)}_{tx}\), and \(E^{i}_{rx}\)}
    \Output{Trained \acrshort{rl} agent/Clustering solution}

    Initialize replay memory \(\mathcal{D}\), Q-network \(Q\), and target network \(Q'\)\;

    \For{episode \(p\) in \(\mathcal{P}\)}{
        Initialize state \(s_0\)\;
        \While{episode \(p\) is not terminated}{
            Select action \(a_t\) based on \(\epsilon\)-greedy strategy\;
            \(r_t \leftarrow 0\)\;
            \Comment{\textcolor{blue}{Execute action \(a_t\)}}
            \If{\(a_t = a_1\)}{
                \If{training mode}{
                    \(CH \leftarrow\) Predict \acrshortpl{ch} using the neural network\;
                    \(CA \leftarrow\) Predict cluster member assignment using the neural network\;
                }
                \Else{
                    \(CH,~CA \leftarrow\) Generate clustering solution using~(\ref{eq:optimization_problem})\;
                }
                \(CH_\tau \leftarrow 0\)\;
                \(r_t \leftarrow r_t + 0.1\)\;
            }
            \Else{
                Maintain the current clustering solution\;
                \(CH_\tau \leftarrow CH_\tau + 1\)\;
            }
            \(r_t \leftarrow r_t + 1\)\;
            Observe next state \(s_{t+1}\)\;
            Store transition \((s_t, a_t, r_t, s_{t+1})\) in \(\mathcal{D}\)\;

            Sample minibatch from \(\mathcal{D}\)\;
            Update Q-network using the sampled transitions\;
            Update target network\;
            \If{training mode}{
                \If{\(\exists n \in N\) such that \(E_n \leq 0\)}{
                    \(r_t \leftarrow r_t + 2\)\;
                    Terminate episode \(p\)\;
                }
            }
            \Else{
                \If{\(|\mathcal{D}| < 1\)}{
                    \Comment{\textcolor{blue}{No more nodes alive}}
                    Terminate episode \(p\)\;
                }
            }
        }
    }
\end{algorithm}

The neural network architecture for the \acrshortpl{ch} is designed to take as input the \(\alpha\), \(\beta\) and \(\gamma\) parameters, \(E_{net}\), \(E_n~\forall n \in N\), \(F\), \(\hat{E}^{ch}_{tx}\), \(\hat{E}^{ch}_{rx}\), and \(\widehat{|CH|}\).
Where \(F=[F_1, F_2, \dots, F_{|N|}]\) is a binary vector indicating whether a node \(n\) is a potential cluster head (\(F_n=1\)) or not.
Node \(n\) is a potential cluster head \(F_n = 1 \iff E_n \geq \bar{E}\), where \(\bar{E}\) is the average energy level of the network.
\(\hat{E}^{ch}_{tx} = [\hat{E}^{ch}_{tx_1}, \hat{E}^{ch}_{tx_2}, \ldots, \hat{E}^{ch}_{tx_{|N|}}]\) is a vector containing the expected energy consumption of potential \acrshortpl{ch} when transmitting data to the \acrshort{bs}.
Here, \(\hat{E}^{ch}_{tx_n} = E^{ch_n}_{tx}\) if \(F_n = 1\), and \(\hat{E}^{ch}_{tx_n} = 0\) otherwise.
The vector \(\hat{E}^{ch}_{rx} = [\hat{E}^{ch}_{rx_1}, \hat{E}^{ch}_{rx_2}, \ldots, \hat{E}^{ch}_{rx_{|N|}}]\) is a representation of the expected energy consumption of potential \acrshortpl{ch} when receiving data from cluster members.
For each potential cluster head \(n\), \(\hat{E}^{ch}_{rx_n}\) equals \(E^{ch_n}_{rx}\) if \(F_n = 1\), and is zero otherwise.
\(\widehat{|CH|} = k\times|\mathcal{D}|\) is the expected number of \acrshortpl{ch}, where \( \mathcal{D} = \{n \in N | E_n > 0\},~\mathcal{D} \subseteq N \) denotes the set of nodes that are currently active or operational, and \(|\mathcal{D}|\) represents the cardinality of \(\mathcal{D}\).
The target output of the neural network is the set of \acrshortpl{ch}.
The module architecture, with a depth of three, consists of three fully connected layers with 401, 2000, and 101 neurons, respectively.
Each layer is followed by a ReLU activation function and a dropout layer.
The output layer produces a set of probabilities using the sigmoid activation function, indicating the likelihood of each node becoming a \acrshort{ch}.
We divide the dataset into training and testing sets, with a ratio of 80:20.
The neural network is trained using the Adam optimizer with a learning rate of $1e-4$ and a batch size of $16$.
The loss function is the binary cross-entropy loss.
The neural network is trained for $1000$ epochs.
The confusion matrix illustrating the performance of the neural network-based \acrshort{ch} prediction is omitted due to space constraints.
However, the neural network achieves an overall accuracy of $99.24\%$ on the testing set.
This outstanding performance is attributed to the utilization of meaningful features as input, which exhibits a high correlation with the \acrshortpl{ch}' feature.

The neural network architecture for the assignment of cluster members is designed to take the following parameters as input: \(\alpha\), \(\beta\), \(\gamma\), \(E_{net}\), \(E_n\) for all \(n \in N\), \(E^{sink}_{tx}\), \(E^{(i,ch_i)}_{tx}\), and \(CH\).
The vector \(E^{sink}_{tx} = [E^{ch_1}_{tx}, E^{ch_2}_{tx}, \ldots, E^{ch_{|CH|}}_{tx}]\) represents the energy consumption of \acrshortpl{ch} when transmitting data to the \acrshort{bs}.
\(|CH|\) represents the cardinality of the set of \acrshortpl{ch}.
If \(|CH| < k \times |N|\), \(E^{sink}_{tx}\) is padded with zeros to match the length of \(CH\).
The vector \(E^{(i,ch_i)}_{tx}\) is defined as \([E^{(i,ch_1)}_{tx}, E^{(i,ch_2)}_{tx}, \ldots, E^{(i,ch_{|CH|})}_{tx}]\) for all \(i \in N\) and \(n \notin CH\).
It represents the energy consumption of cluster members when transmitting data to each \acrshort{ch}.
Lastly, \(CH\) is the set of cluster heads, denoted as \(CH = \{CH_1, CH_2, \ldots, CH_{|CH|}\}\) taken from the output of the neural network-based cluster head prediction module.
The target output of the neural network is the assignment of cluster members to each \(ch_i \in CH\).
The module architecture, with a depth of three, consists of three fully connected layers with $703$, $2000$, and  $|CH_{max}|\times|N|$ neurons, respectively.
Where $|CH_{max}|=k\times|N|$ is the maximum number of \acrshortpl{ch}.
The output layer is reshaped to a $(|N|, |CH_{max}|)$ matrix, where each row represents the probability distribution of a node being assigned to each \(ch_i \in CH\).
Each layer is followed by a ReLU activation function and a dropout layer.
The output layer produces a set of probability distributions using the softmax activation function, indicating the likelihood of each node being assigned to a cluster head.
The dataset is also divided into training and testing sets, with a ratio of 80:20, and the neural network is trained using the Adam optimizer with a learning rate of $1e-6$ and a batch size of $16$.
The loss function is the categorical cross-entropy loss.
The neural network is trained for $1000$ epochs.
Fig.~\ref{fig:ca_confusion_matrix} shows the confusion matrix, offering insights into the performance of the neural network-based cluster member assignment.
It is noteworthy that the model predicts the cluster head index for each node based on the cluster head set.
Row and column accuracies are also provided on the right and top sides of the confusion matrix, respectively.
The neural network achieves an overall accuracy of $96.74\%$ on the testing set.

\subsection{Training RL agent and surrogate model}
\label{subsubsec:training_rl_agent_and_surrogate_model}

We use the \acrfull{dqn} algorithm to train the \acrshort{rl} agent~\cite{hesterDeepQlearningDemonstrations2018}.
This algorithm is inspired by the Q-Learning algorithm, which is a model-free \acrshort{rl} algorithm.
It mainly comprises deep neural networks to approximate the action-value function, a replay memory to store the agent's experiences, and a target network to stabilize the learning process.
It updates the state-action value function, also known as the Q-function, using the Bellman equation~\cite{watkinsQlearning1992}.
The training process of \acrshort{leach-rlc} is summarized in Algorithm~\ref{alg:leach-rlc}.
The surrogate model to predict the clustering solution is only used during the training process.
For the evaluation, we use the \acrshort{milp} to generate the new clusters as it provides the optimal solution.

For training purposes, we set (\(\alpha, \beta, \gamma\)) to ($54.83$, $14.54$, $35.31$), which are the optimal values obtained from the \acrshort{milp} formulation.
We train the \acrshort{rl} agent with the following parameters: a learning rate of $1e-4$, a discount factor of $0.90$, an exploration rate of $0.8$, a batch size of $128$, and a target network update frequency of $100$.
The agent is trained for $200k$ time steps.

It is important to note that online training of the \acrshort{rl} agent is not feasible due to the large number of episodes required, as well as the extensive exploration and parameter tuning involved.
Such training would be impractical in real-world WSN deployments, where decision intervals typically range from seconds to minutes in some monitoring applications.
The long operational time required to observe agent learning and adjust parameters makes real-time training infeasible.
Therefore, we have opted for offline training, which allows us to train the agent in a controlled environment and thoroughly evaluate its performance across various scenarios.
After training, the agent can be deployed in real-world scenarios, where it can make decisions based on the learned policy.

\subsubsection{Energy consumption during training}

While energy consumption is critical in real-world \acrshort{wsn}, it is important to highlight that training of the \acrshort{rl} agent is conducted entirely offline in a simulation-based environment.
This design ensures that training energy consumption has no direct impact on real-world deployments.
Offline training allows us to thoroughly evaluate various configurations and optimize the model.
During deployment, the energy overhead is limited to inference and signaling updates, which are computationally efficient and have been evaluated in Section~\ref{sec:results}.

\subsubsection{Training convergence}

\begin{figure}[!htbp]
    \centering
    \includegraphics[width=0.98\columnwidth]{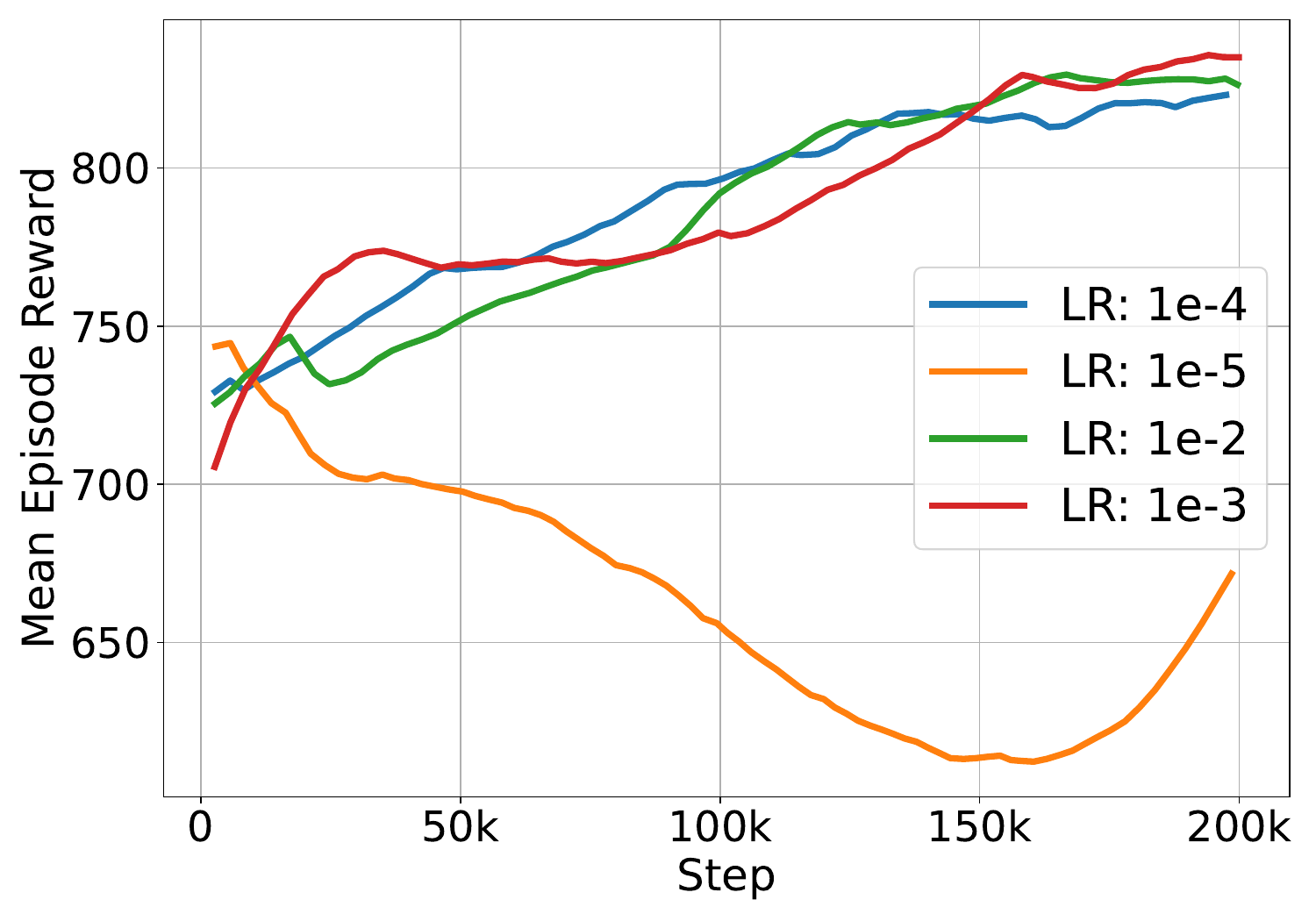}
    \caption{Convergence behavior of the RL agent for different learning rates.}
    \label{fig:convergence}
\end{figure}

To ensure the effectiveness of the \acrshort{rl} agent, we monitored the convergence of the training process by varying critical parameters such as the learning rate ($1 \times 10^{-2}$, $1 \times 10^{-3}$, $1 \times 10^{-4}$, and $1 \times 10^{-5}$).
The results, shown in Fig.~\ref{fig:convergence}, reveal that learning rates of $1 \times 10^{-2}$, $1 \times 10^{-3}$, and $1 \times 10^{-4}$ exhibited similar behavior, converging to a comparable mean episode reward.
However, the learning rate of $1 \times 10^{-5}$ struggled to converge, indicating insufficient progress during training.
Among the tested rates, $1 \times 10^{-4}$ provided a slightly more stable convergence than the higher learning rates, striking a balance between convergence speed and stability.
This analysis led us to choose $1 \times 10^{-4}$ as the optimal learning rate for the training process.

Although the number of training steps involved was $200k$, it is worth mentioning that this comprised approximately $270$ episodes.
On average, each episode lasted $740$ steps before the first node died.
Naturally, some episodes were shorter, particularly during the early stages of training, as the agent had yet to learn when to conserve energy or generate a new cluster.
These shorter episodes reflect the exploratory nature of the agent during the initial phases of training.

\begin{table}[!htbp]
    \centering
    \caption{Network parameters.}
    \label{tab:network_parameters}
    \begin{NiceTabular}[c]{ll}[hvlines]
        \CodeBefore
        \rowcolor{lightgray}{1}
        \rowcolors{2}{gray!12}{}[respect-blocks]
        \Body
        \RowStyle[]{\bfseries}
        Parameter              & Value                              \\
        \(N\)                  & 100                                \\
        \(L\)                  & 100 m                              \\
        \acrshort{bs} location & (50, 175)                          \\
        \(E_0\)                & 0.5 J                              \\
        \(E_{elec}\)           & 50 nJ/bit                          \\
        \(E_{fs}\)             & 10 pJ/bit/m\textsuperscript{2}     \\
        \(E_{amp}\)            & 0.0013 pJ/bit/m\textsuperscript{4} \\
        \(F_{DA}\)             & 5 nJ/bit                           \\
        \(B\)                  & 4000 bits                          \\
        \(B_c\)                & 1000 bits                          \\
        \(k\)                  & 0.05                               \\
    \end{NiceTabular}
\end{table}

\section{Results}
\label{sec:results}

In this section, we assess the performance of \acrshort{leach-rlc} by conducting a comparative analysis against its counterparts.
The proposed clustering protocol and the \acrshort{rl} agent are implemented using Python.
The \acrfull{glpk} solver~\cite{makhorinGLPKGNULinear2012} is employed to solve the \acrshort{milp}, while the Stable Baselines3 library~\cite{raffinStablebaselines3ReliableReinforcement2021} is utilized for implementing the \acrshort{rl} agent.

\begin{figure*}[!htbp]
    \centering
    \subfloat[\label{fig:alive_nodes_vs_rounds}]{%
        \includegraphics[width=0.98\columnwidth]{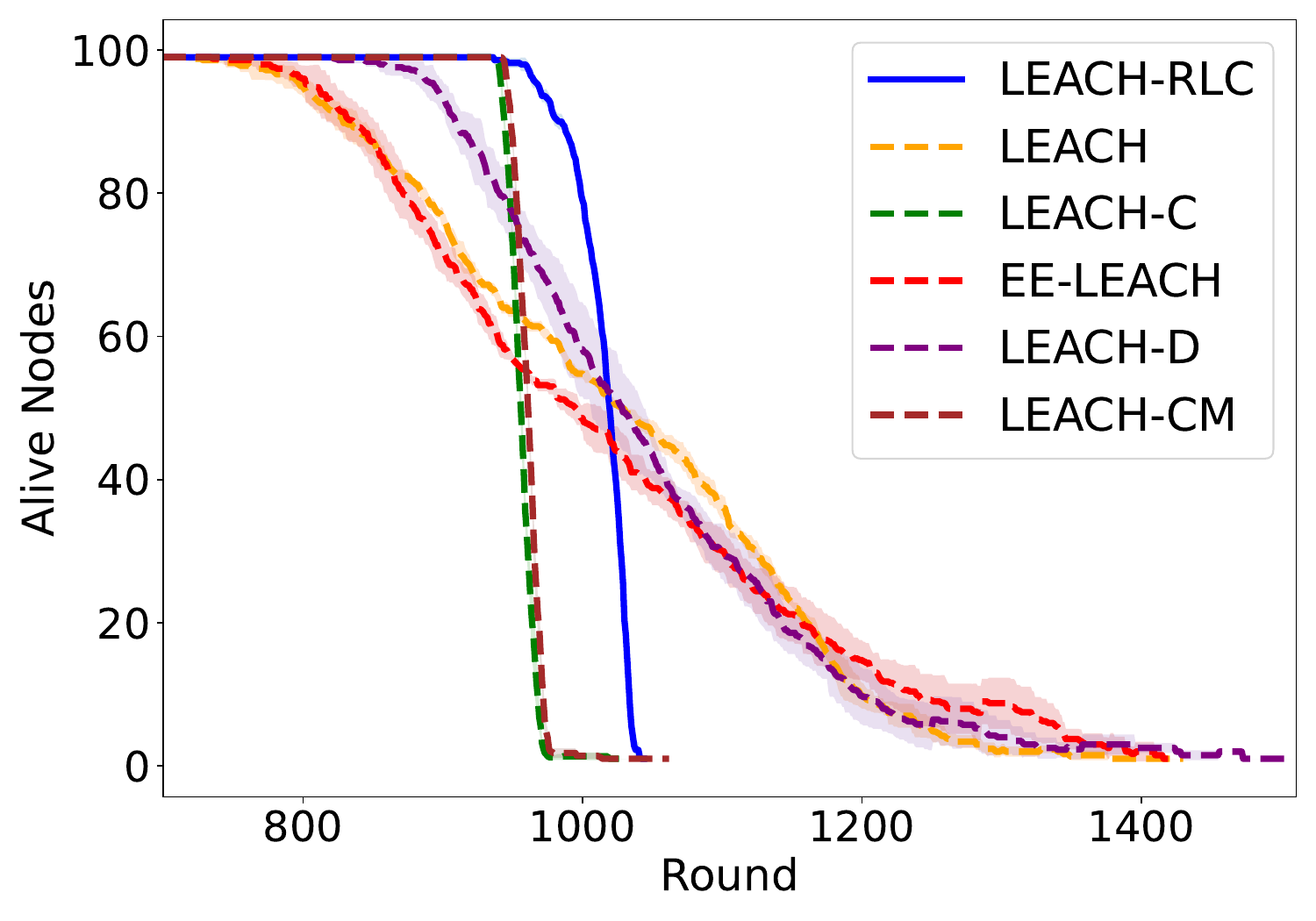}
    }
    \subfloat[\label{fig:fnd}]{%
        \includegraphics[width=0.98\columnwidth]{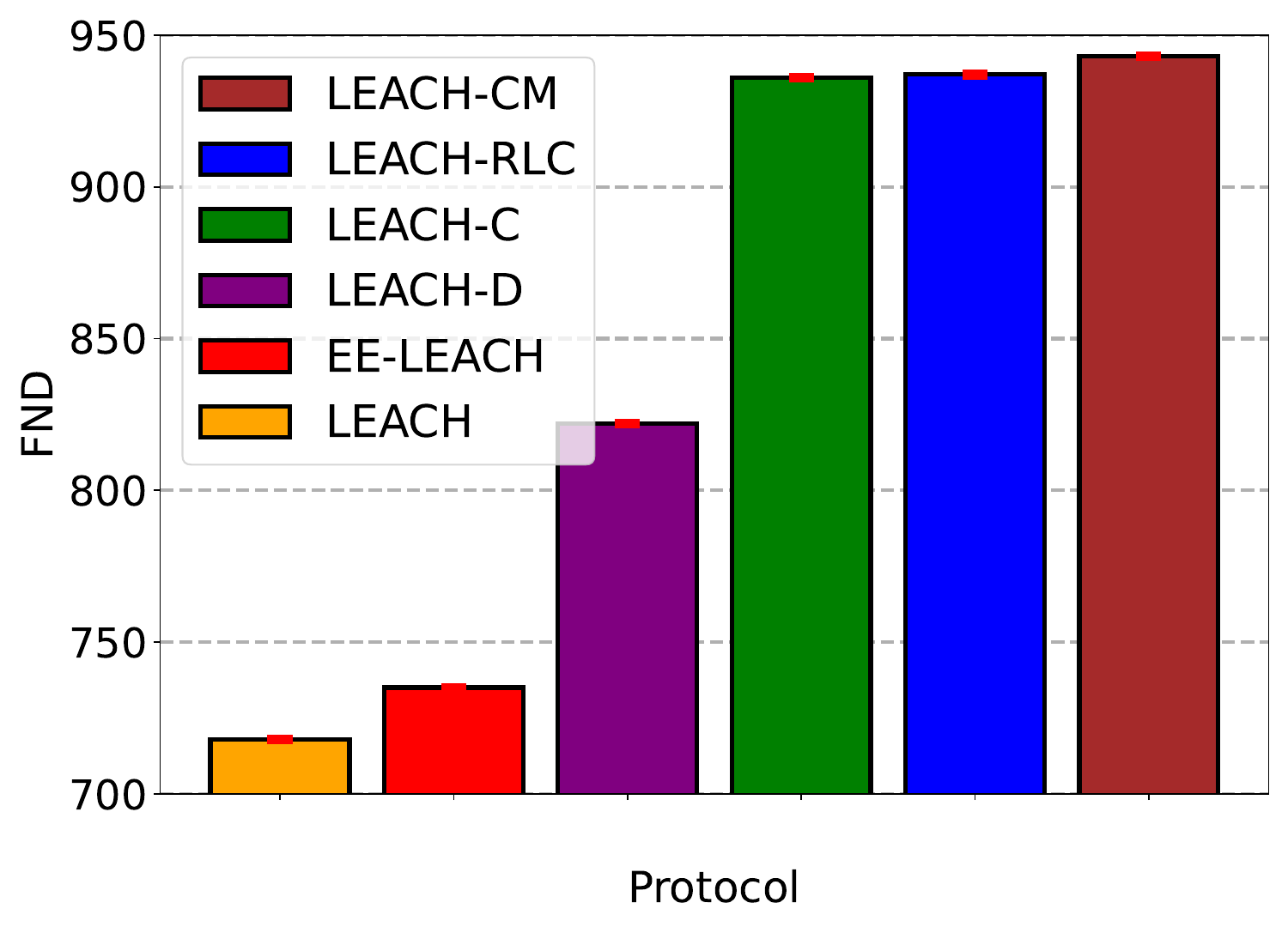}
    }
    \caption{Network lifetime. (a) Alive nodes vs. rounds. (b) First node dead.}
    \label{fig:network_lifetime}
\end{figure*}

\begin{figure}[!htbp]
    \centering
    \includegraphics[width=0.98\columnwidth]{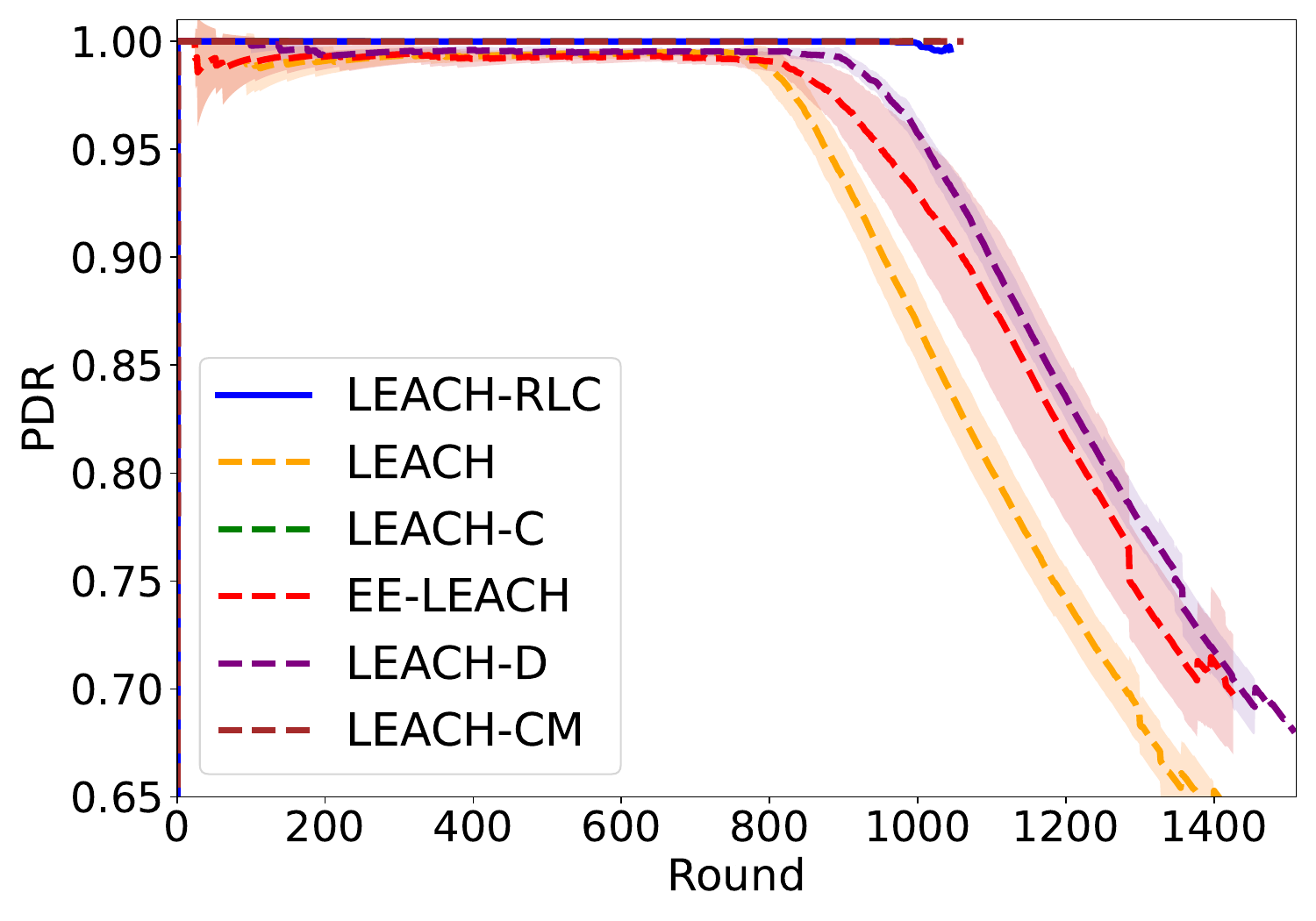}
    \caption{\acrfull{pdr}.}
    \label{fig:pdr}
\end{figure}

\begin{figure*}[!htbp]
    \centering
    \subfloat[\label{fig:average_energy_consumption}]{%
        \includegraphics[width=0.98\columnwidth]{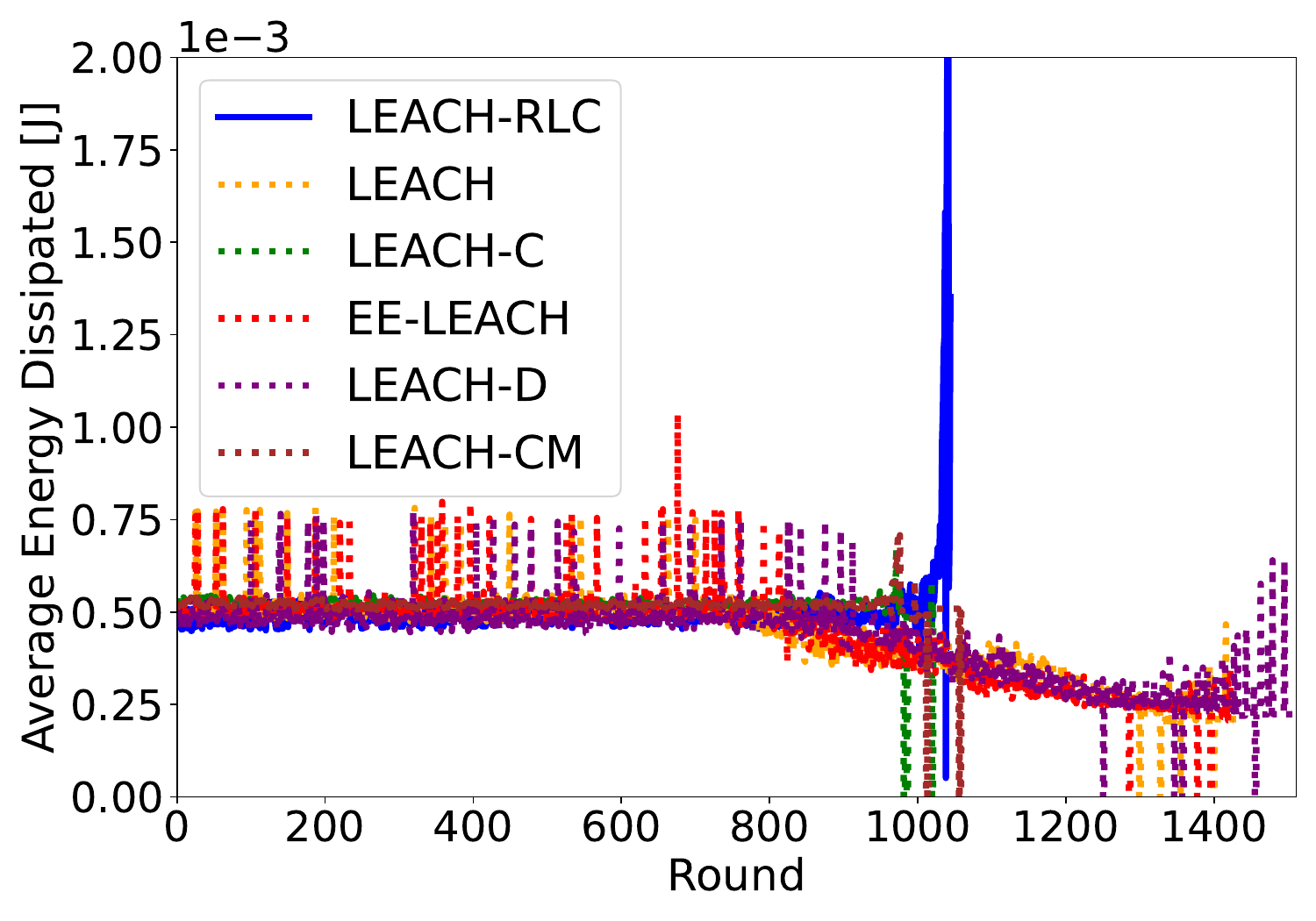}
    }
    \subfloat[\label{fig:remaining_energy}]{%
        \includegraphics[width=0.98\columnwidth]{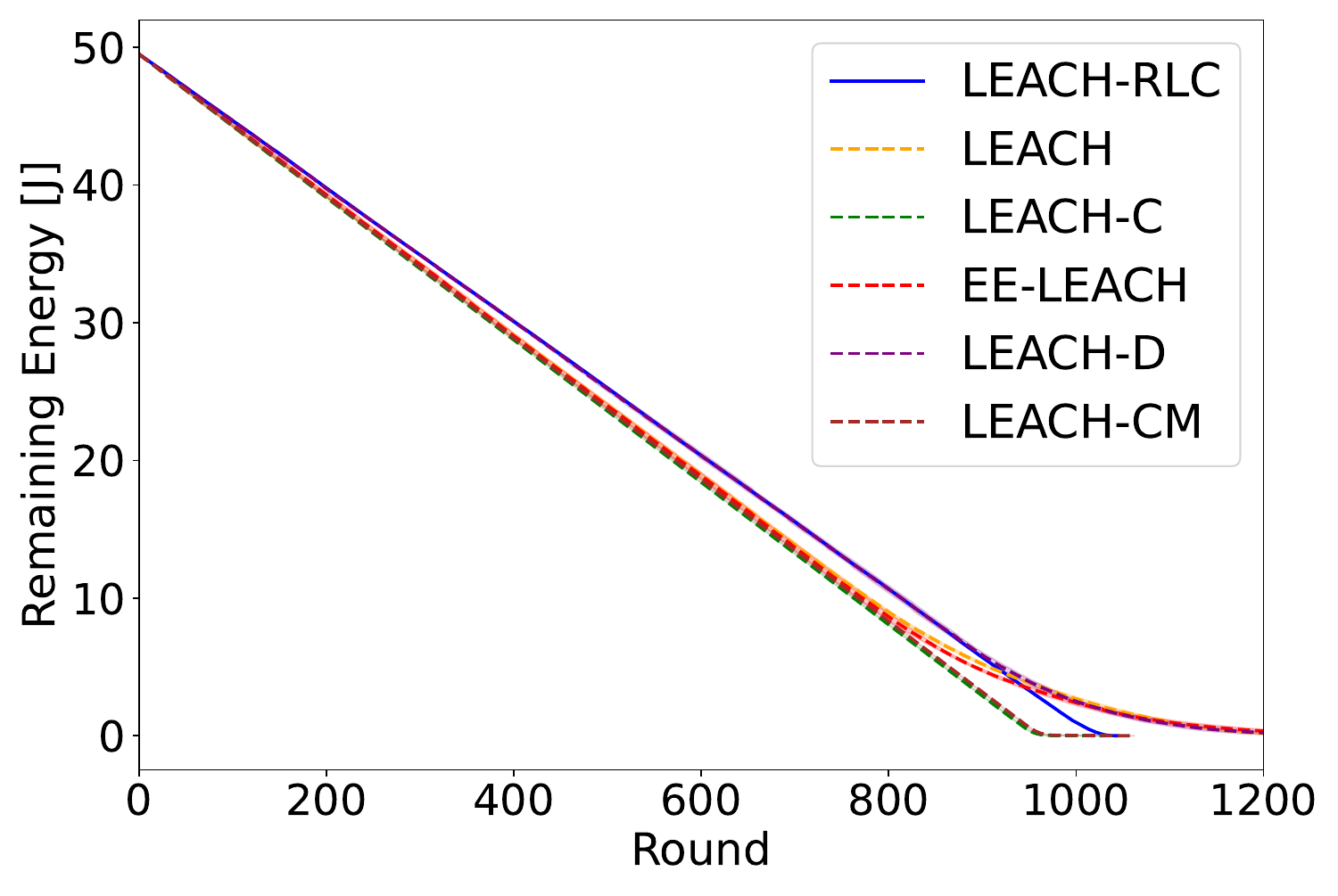}
    }
    \caption{Average network energy dissipated (a) and remaining energy of the network (b).}
    \label{fig:energy_consumption}
\end{figure*}

\begin{figure*}[!htbp]
    \centering
    \subfloat[\label{fig:leach_rlc_ch_heatmap}]{%
        \includegraphics[width=0.32\textwidth]{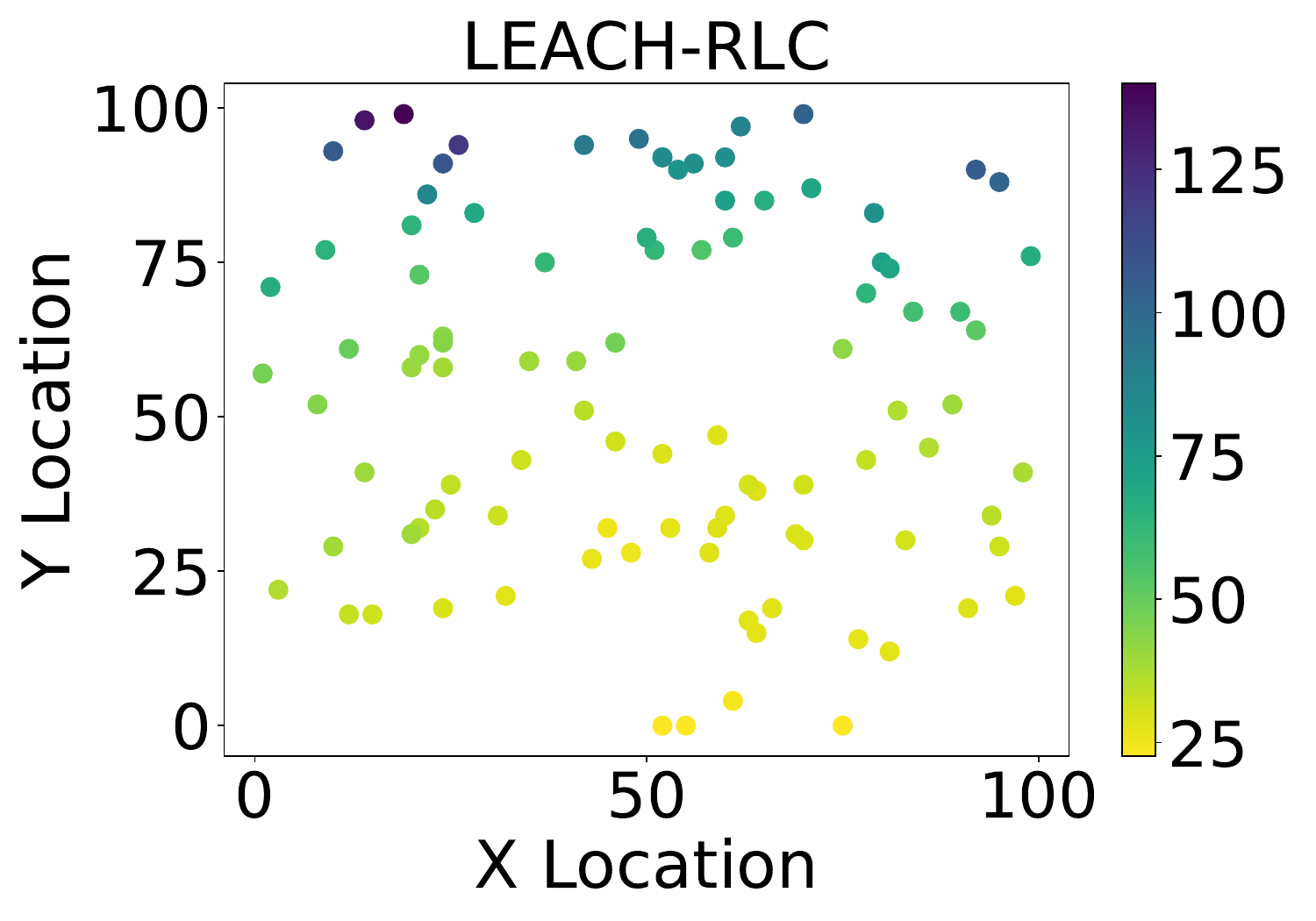}
    }
    \subfloat[\label{fig:leach_heatmap}]{%
        \includegraphics[width=0.32\textwidth]{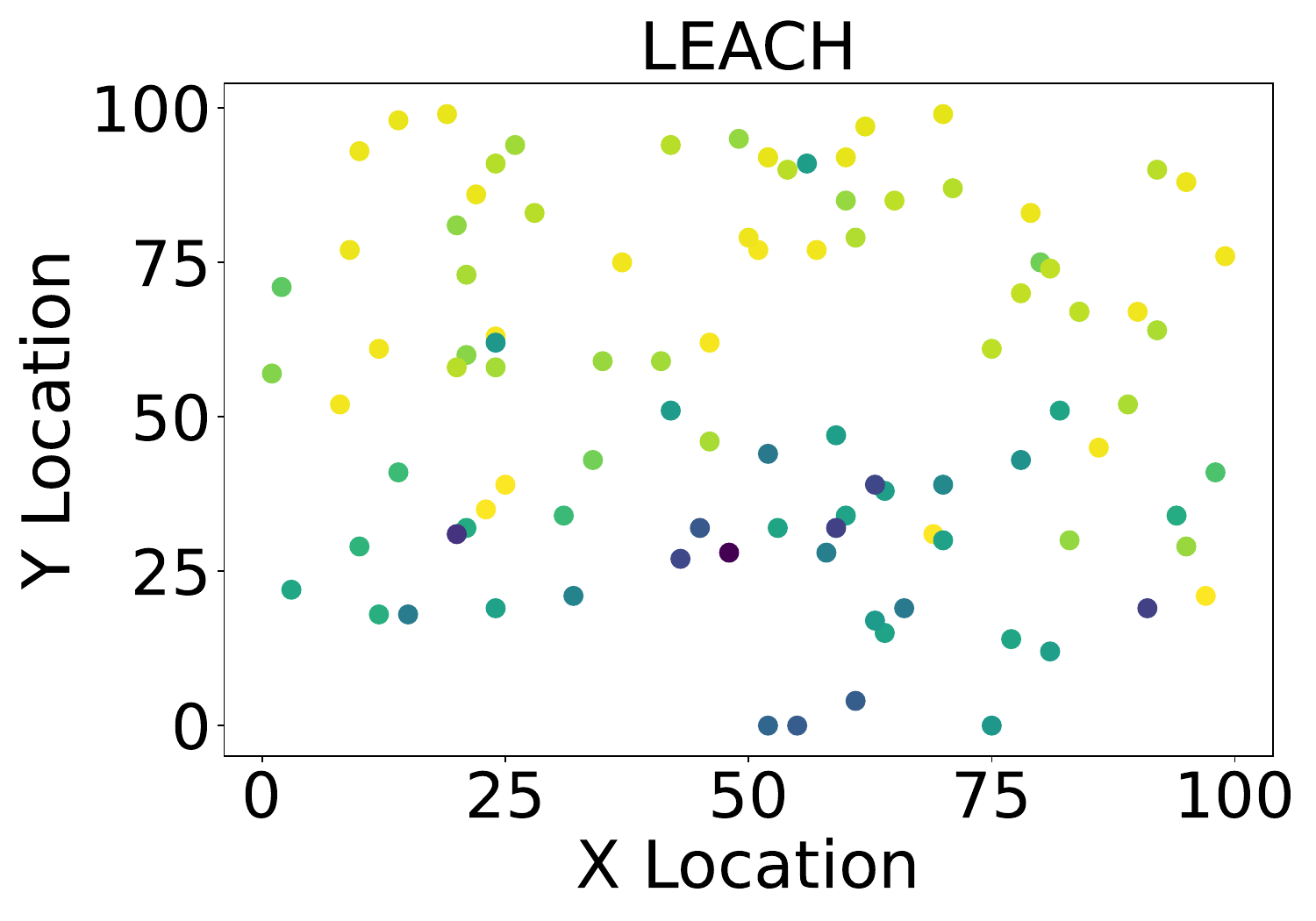}
    }
    \subfloat[\label{fig:leach_c_ch_heatmap}]{%
        \includegraphics[width=0.32\textwidth]{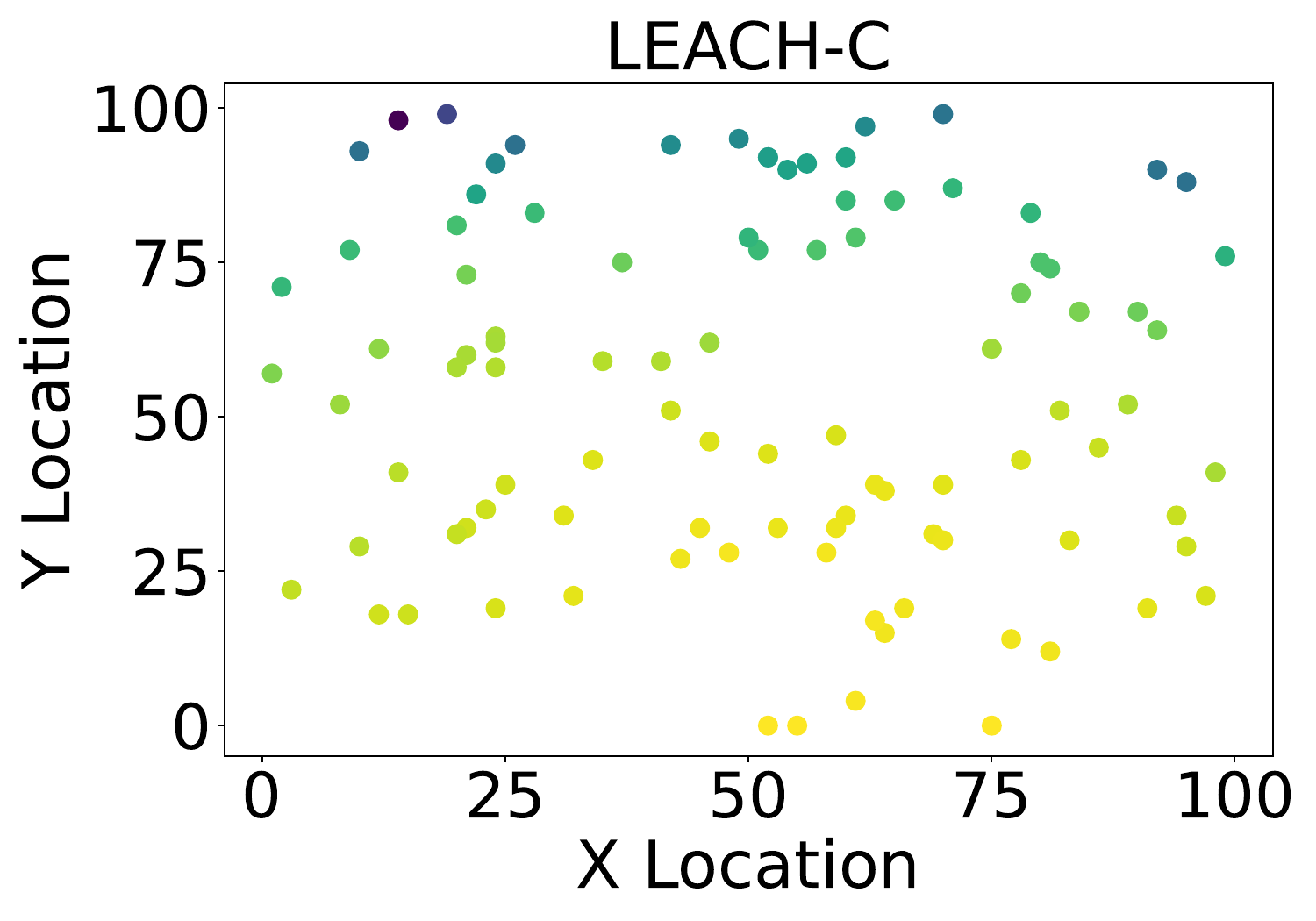}
    }
    \vspace{0.05in} %
    \subfloat[\label{fig:ee_leach_heatmap}]{%
        \includegraphics[width=0.32\textwidth]{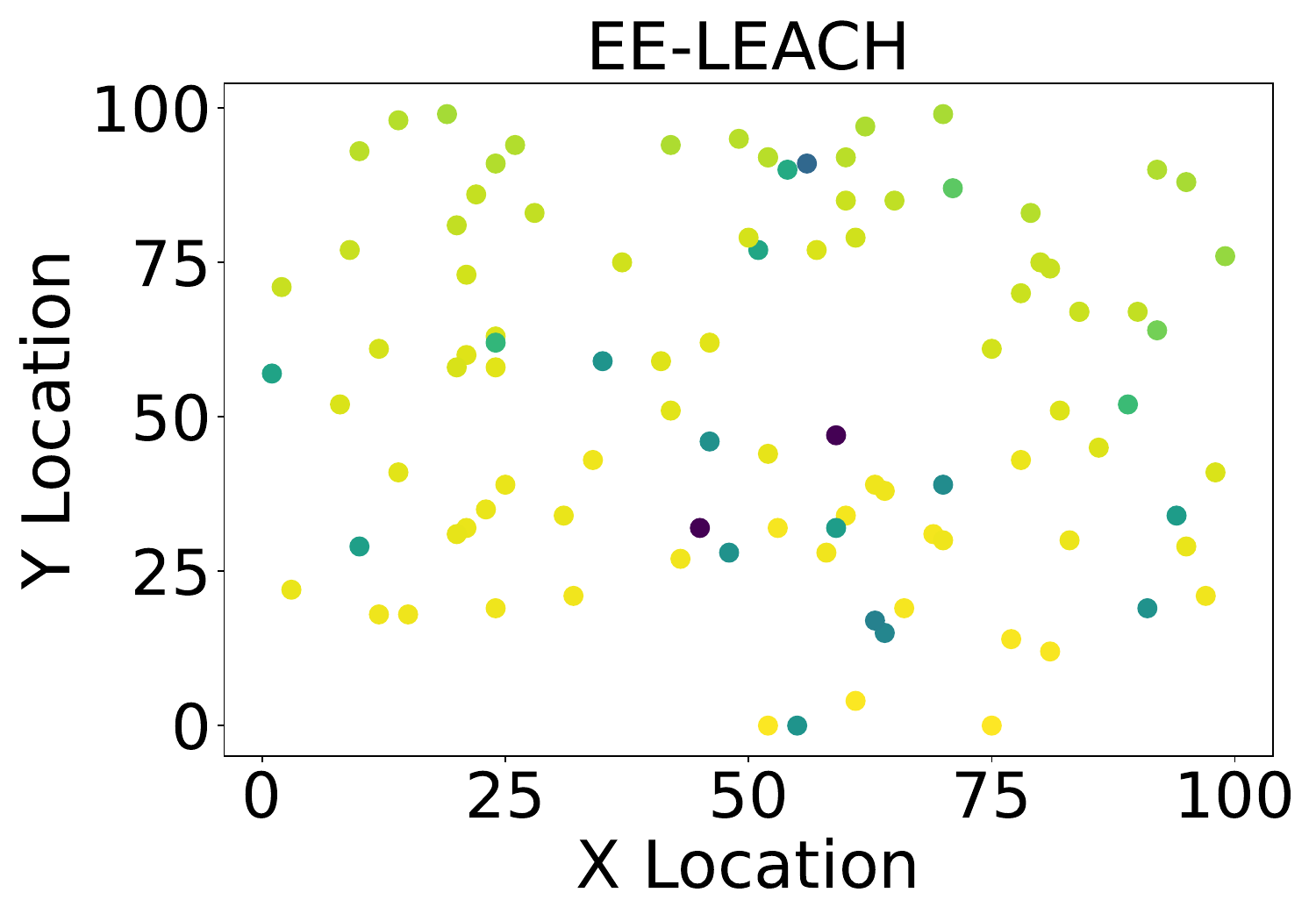}
    }
    \subfloat[\label{fig:leach_d_heatmap}]{%
        \includegraphics[width=0.32\textwidth]{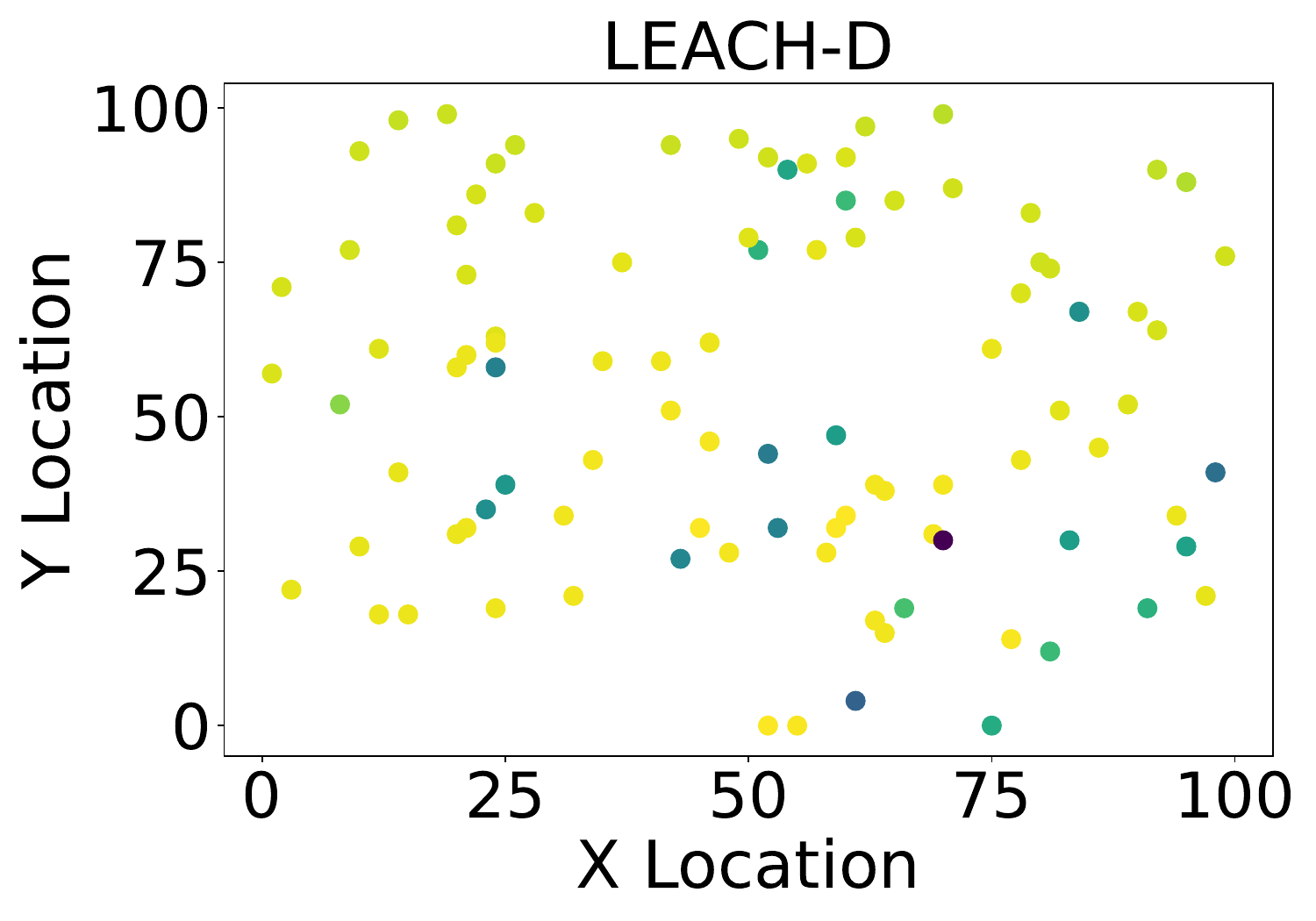}
    }
    \subfloat[\label{fig:leach_cm_heatmap}]{%
        \includegraphics[width=0.32\textwidth]{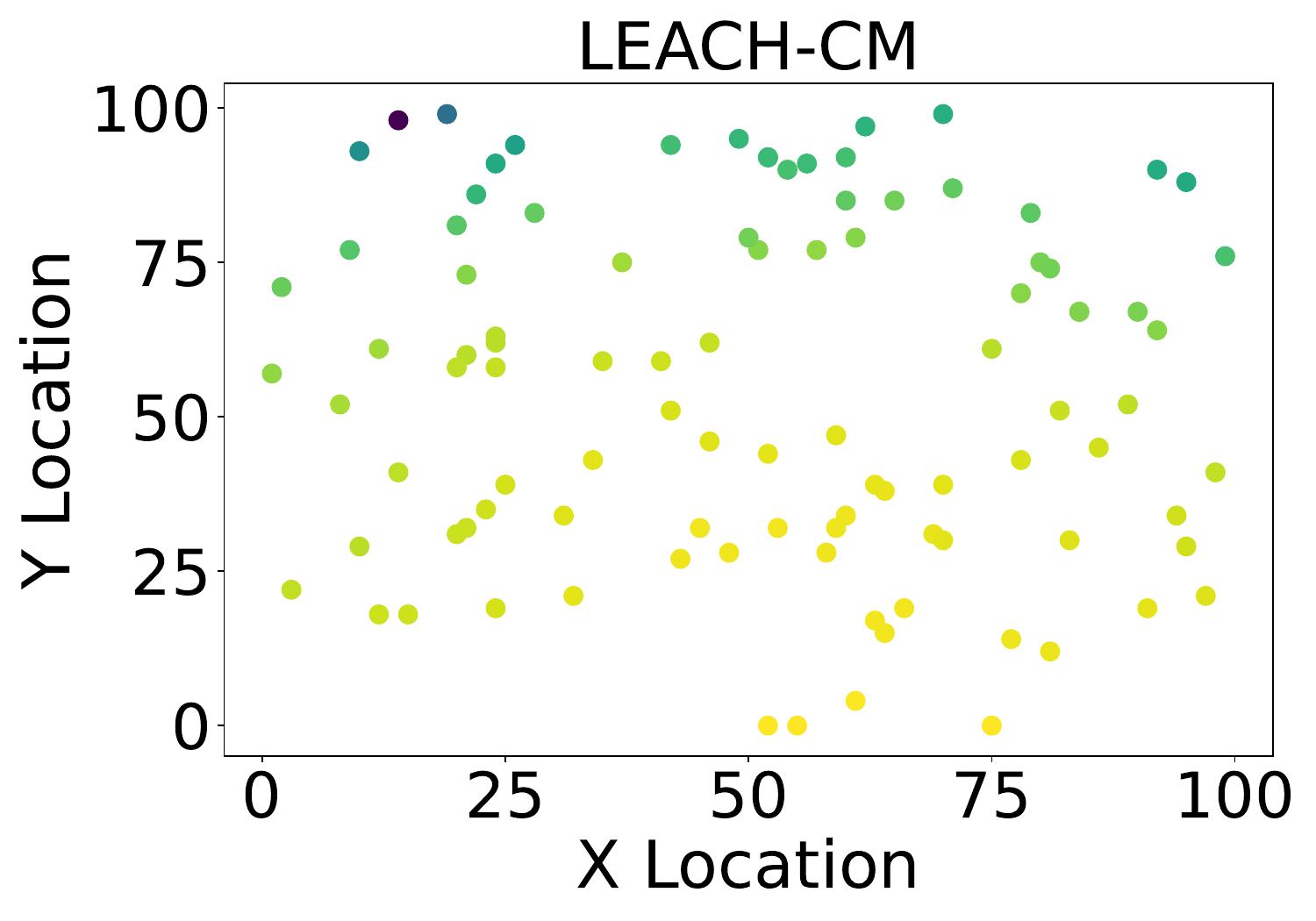}
    }
    \caption{Heatmaps illustrating the frequency of node selection as a cluster head across different protocols. Panels (a) to (f) display heatmaps for LEACH-RLC, LEACH, LEACH-C, EE-LEACH, LEACH-D, and LEACH-CM, respectively.}
    \label{fig:ch_heatmaps}
\end{figure*}

\begin{figure*}[!ht]
    \centering
    \subfloat[\label{fig:leach_rlc_histogram}]{%
        \includegraphics[width=0.16\textwidth]{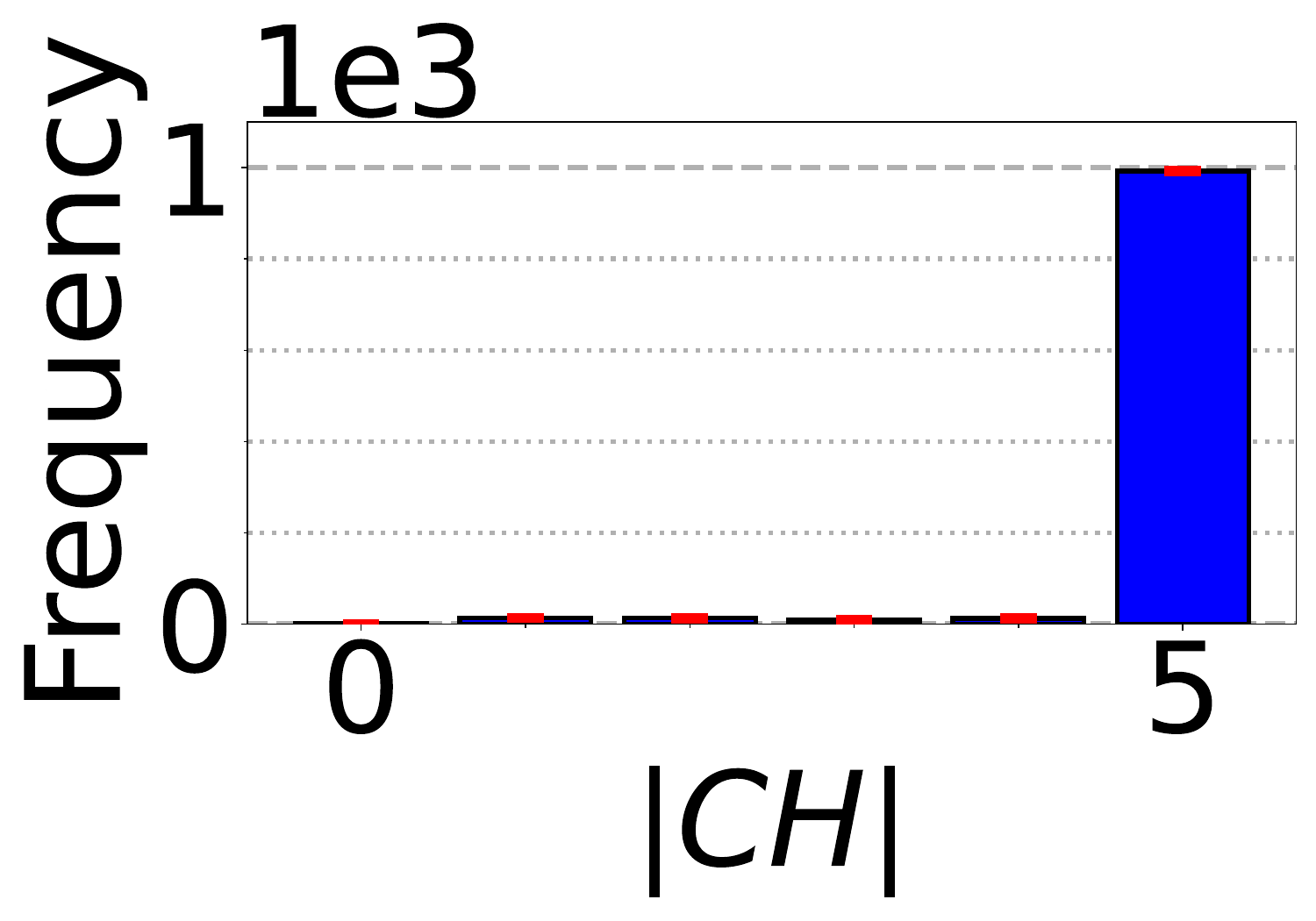}
    }
    \subfloat[\label{fig:leach_histogram}]{%
        \includegraphics[width=0.16\textwidth]{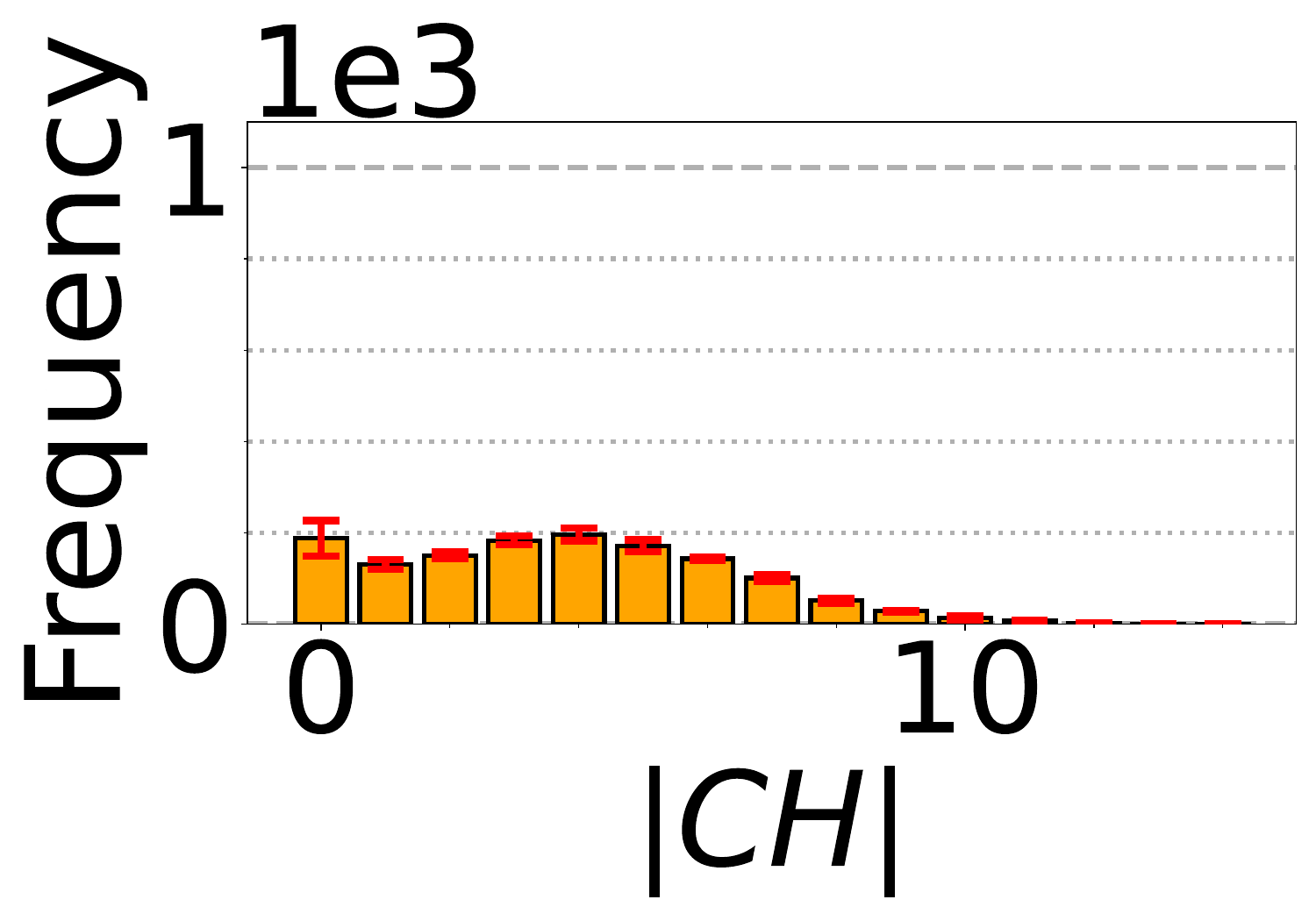}
    }
    \subfloat[\label{fig:leach_c_histogram}]{%
        \includegraphics[width=0.16\textwidth]{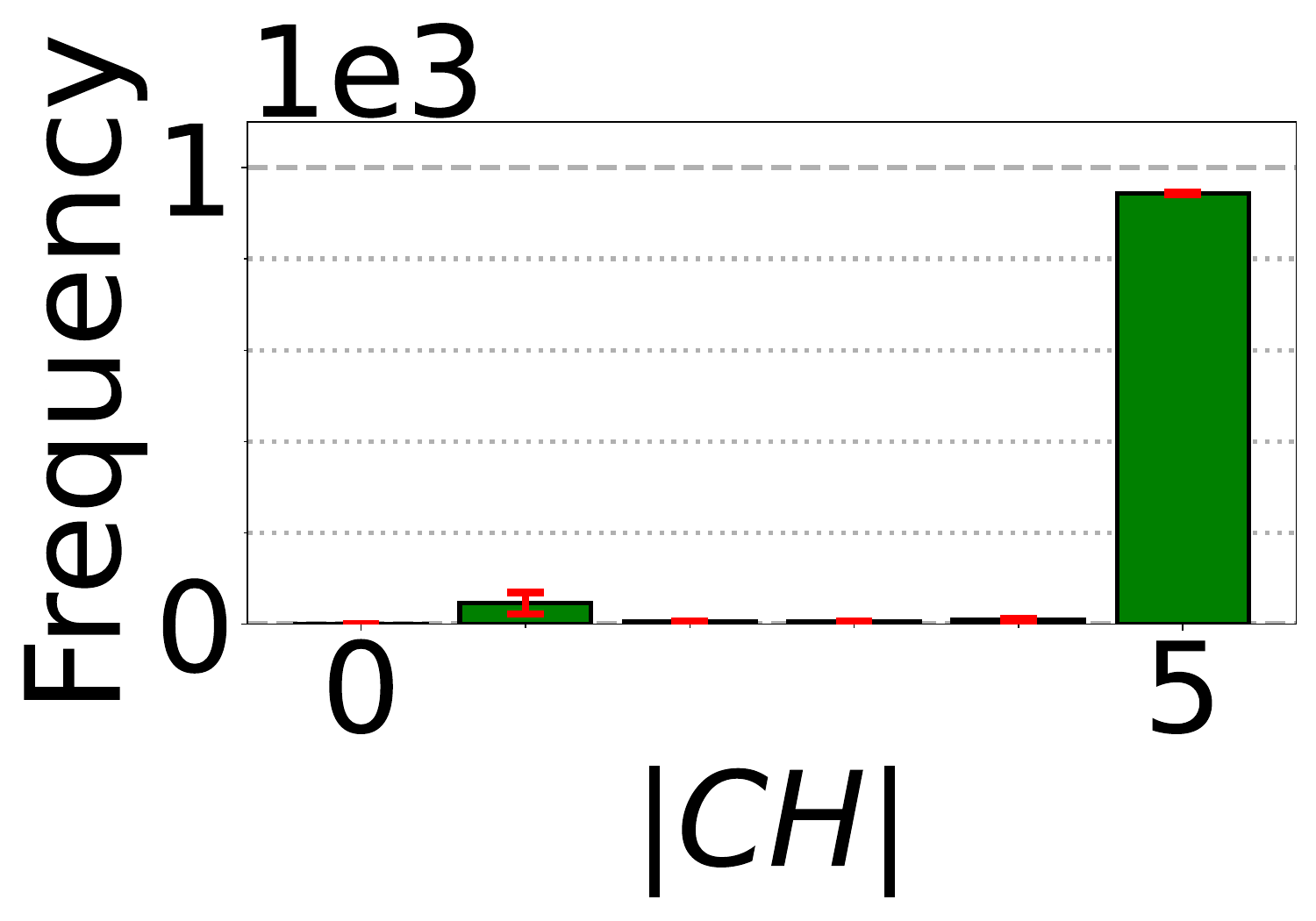}
    }
    \subfloat[\label{fig:ee_leach_histogram}]{%
        \includegraphics[width=0.16\textwidth]{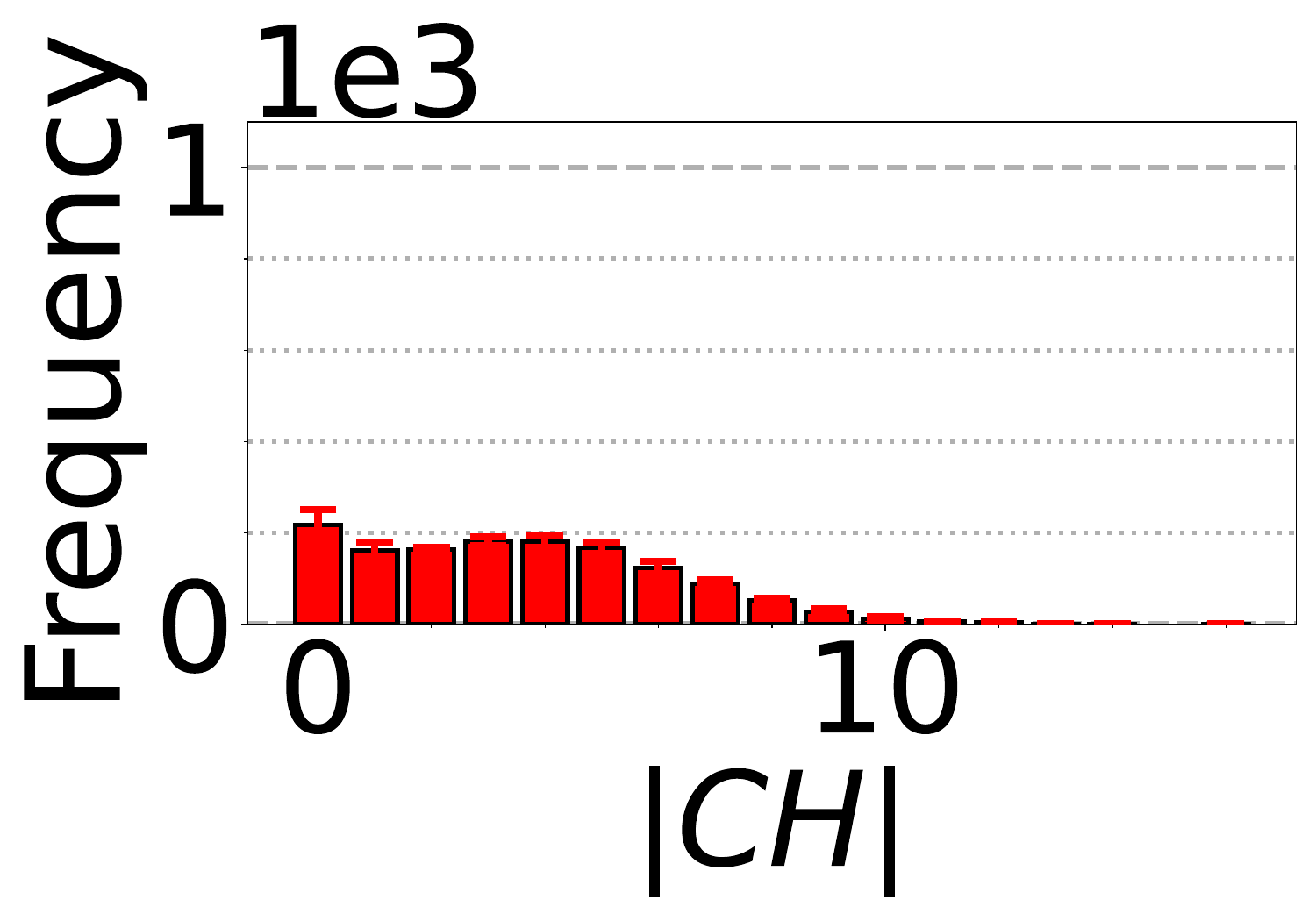}
    }
    \subfloat[\label{fig:leach_d_histogram}]{%
        \includegraphics[width=0.16\textwidth]{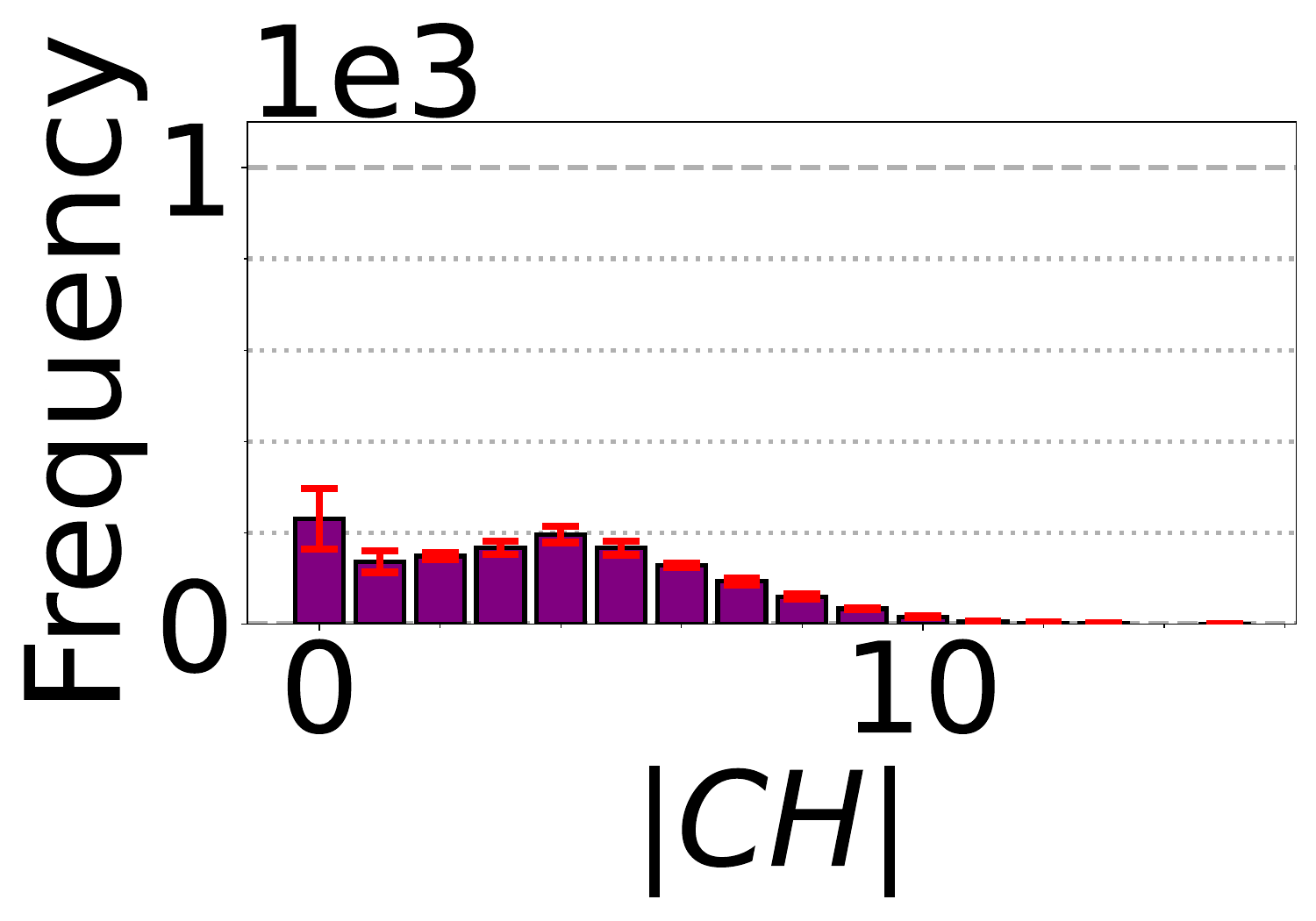}
    }
    \subfloat[\label{fig:leach_cm_histogram}]{%
        \includegraphics[width=0.16\textwidth]{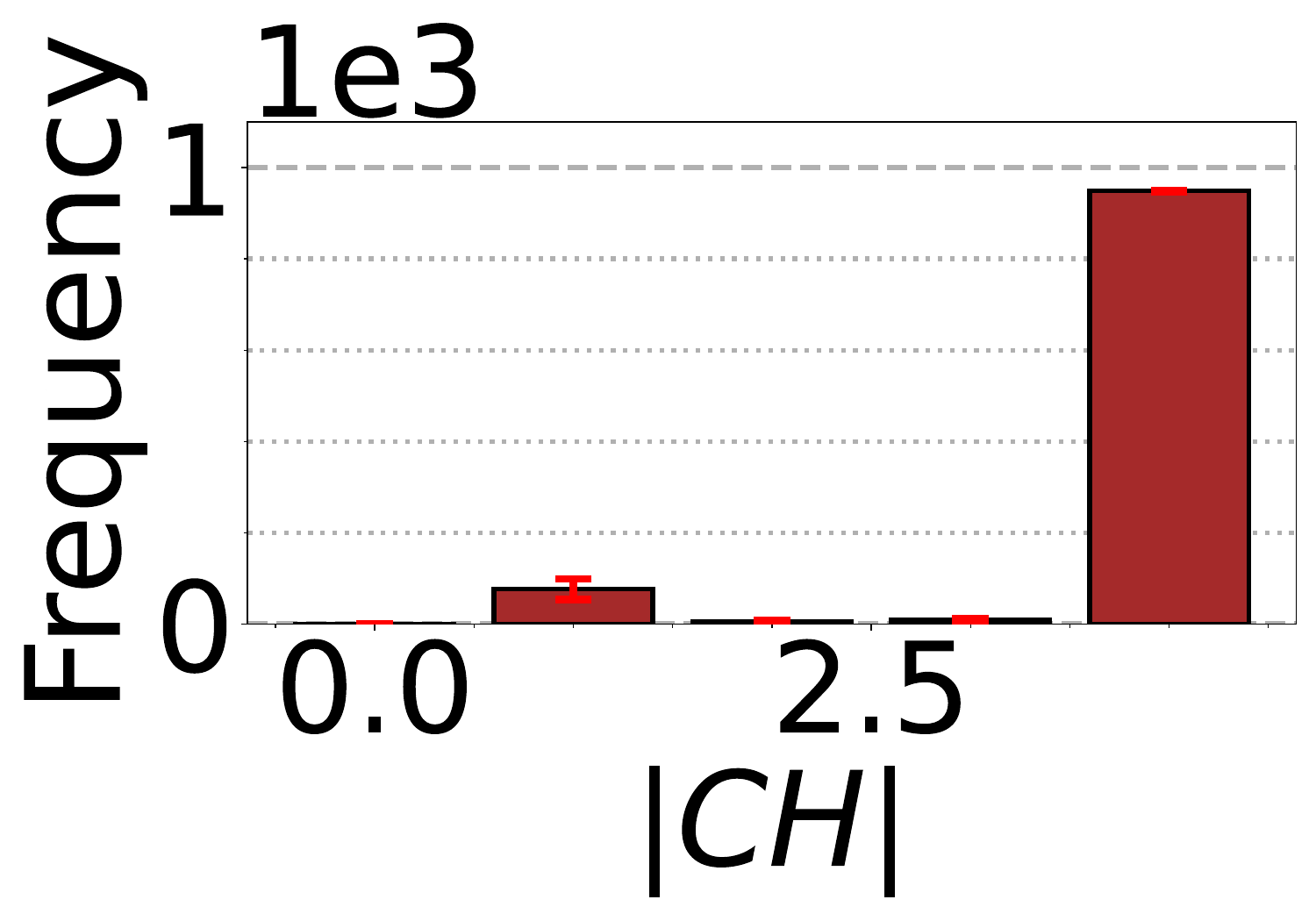}
    }
    \vspace{0.05in} %
    \caption{Histograms depicting the distribution of cluster head selections for each protocol. Panels (a) to (f) correspond to LEACH-RLC, LEACH, LEACH-C, EE-LEACH, LEACH-D, and LEACH-CM, respectively.}
    \label{fig:histograms}
\end{figure*}

\begin{figure}[!htbp]
    \centering
    \includegraphics[width=0.98\columnwidth]{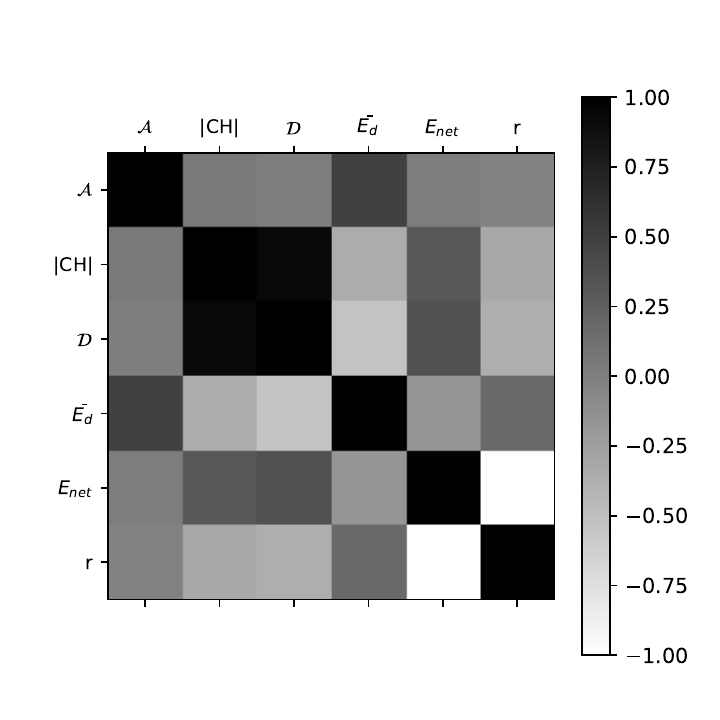}
    \caption{Correlation matrix for the \acrshort{rl} agent.}
    \label{fig:correlation_matrix_rl}
\end{figure}

\begin{figure}[!htbp]
    \centering
    \includegraphics[width=0.98\columnwidth]{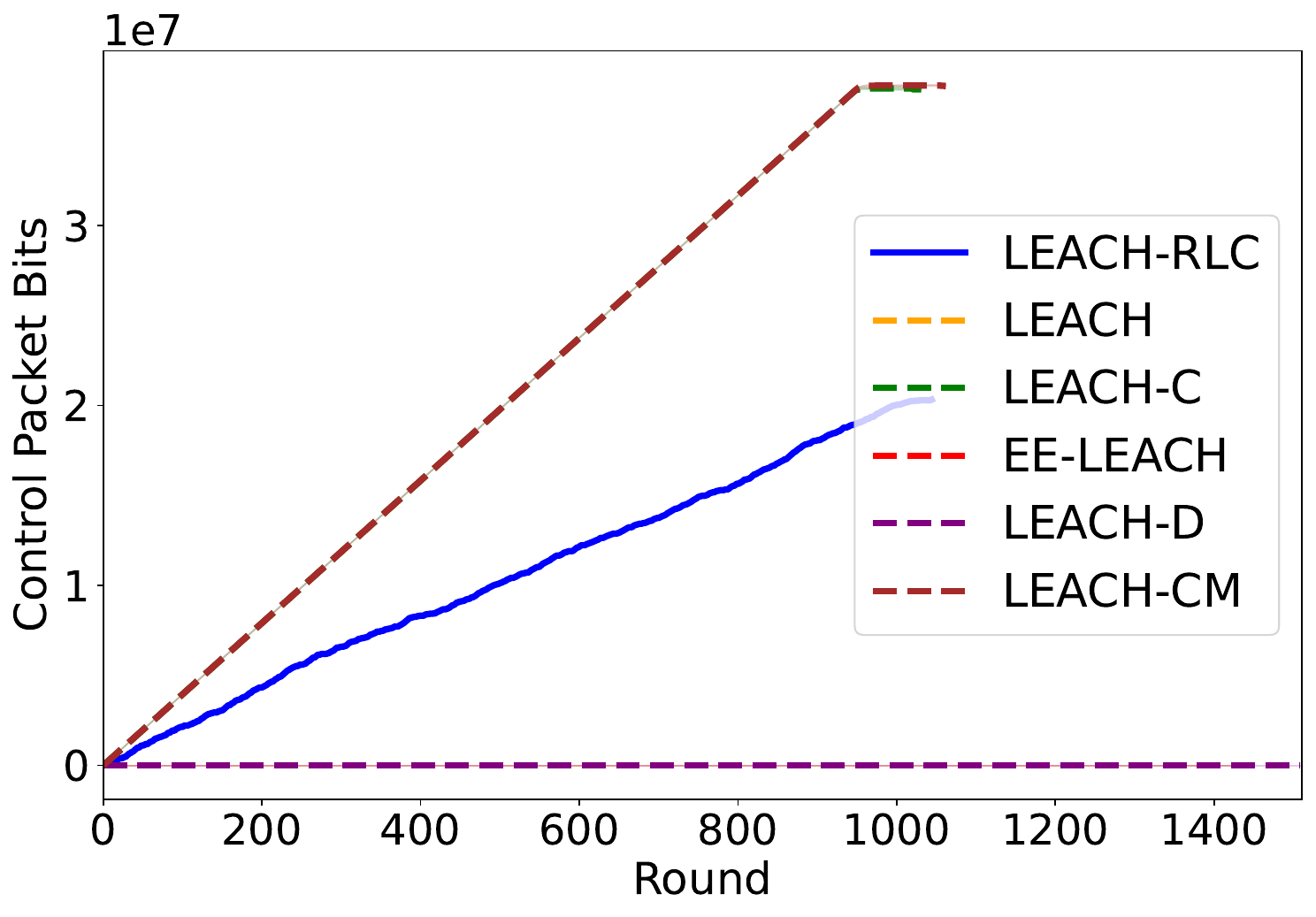}
    \caption{Control overhead.}
    \label{fig:control_overhead}
\end{figure}

\subsection{Simulation Setup}

The network parameters used in the simulations are outlined in Table~\ref{tab:network_parameters}.
Nodes are randomly deployed within a 100\,m\,$\times$\,100\,m area, with the \acrshort{bs} strategically positioned at 50\,m, 175\,m.
This setup ensures that communication between the nodes and the \acrshort{bs} occurs through the intermediary \acrshortpl{ch}.

The performance evaluation of the protocols encompasses metrics such as network lifetime, average network energy consumption, \acrshort{ch} selection, and control overhead.

\subsection{Baseline Protocols}
We compare the performance of \acrshort{leach-rlc} against five baseline protocols:

\begin{enumerate}
    \item We consider \acrshort{leach}~\cite{heinzelmanEnergyefficientCommunicationProtocol2002} as the baseline protocol, which is a self-organizing protocol that allows nodes to form clusters without a centralized controller.
    \item We also consider \acrshort{leach-c}~\cite{heinzelmanApplicationspecificProtocolArchitecture2002a}, which uses a centralized controller for \acrshort{ch} selection.
    \item We include EE-LEACH~\cite{bharanyEnergyEfficientClusteringScheme2021a}, which optimizes the clustering threshold equation by considering additional factors like energy optimization.
    \item We consider LEACH-D~\cite{liuLEACHDLowenergyLowdelay2024a}, which optimizes the clustering threshold equation by considering additional factors like drainage rate.
    \item We include LEACH-CM~\cite{parmarImprovedModifiedLEACHC2016}, which is an enhanced version of \acrshort{leach-c}.
\end{enumerate}
For each protocol, we run five simulations and use a 95\% confidence interval to ensure the reliability of the results.

\subsection{Performance Evaluation}

Fig.~\ref{fig:network_lifetime} offers a comprehensive evaluation of network lifetime, measured by the number of rounds until the occurrence of the \acrshort{fnd} event.
In Fig.~\ref{fig:alive_nodes_vs_rounds}, we present the behavior of alive nodes across rounds, while Fig.~\ref{fig:fnd} specifically illustrates the progression leading to the \acrshort{fnd}.

The results from both figures highlight that \acrshort{leach-rlc} significantly outperforms its counterparts, maintaining the highest number of nodes alive for a longer period.
Additionally, \acrshort{leach-rlc} exceeds \acrshort{leach-c} in the \acrfull{hnd} metric, achieving approximately 1050 rounds compared to \acrshort{leach-c} and LEACH-CM, which achieve approximately 950 rounds.
This improvement can be attributed to \acrshort{leach-rlc}'s ability to select \acrshortpl{ch} and allocate nodes to clusters in a more balanced manner, as depicted in Fig.~\ref{fig:ch_heatmaps}, which illustrates the frequency of node selection as \acrshortpl{ch}.

Fig.~\ref{fig:alive_nodes_vs_rounds} further demonstrates that the proposed \acrshort{leach-rlc} algorithm balances energy consumption across all nodes, ensuring fairness in workload distribution.
This leads to most nodes depleting their energy within a close range of rounds, as opposed to other protocols where uneven energy usage results in some nodes dying early while others survive longer.
While this may create the impression that other protocols sustain more nodes over time, this comes at the cost of reduced network coverage, unbalanced workload distribution, and inefficient network lifetime metrics.
\acrshort{leach-rlc} avoids these issues by ensuring all nodes contribute equitably, maintaining consistent network performance and operational fairness, which are critical for energy-constrained networks.

While \acrshort{leach}, EE-LEACH and LEACH-D (Fig.~\ref{fig:leach_heatmap}, Fig.~\ref{fig:ee_leach_heatmap}, and Fig.~\ref{fig:leach_d_heatmap}) tend to select scattered nodes across the network, \acrshort{leach-rlc}, \acrshort{leach-c}, and LEACH-CM (Fig.~\ref{fig:leach_rlc_ch_heatmap}, Fig.~\ref{fig:leach_c_ch_heatmap}, and Fig.~\ref{fig:leach_cm_heatmap}) prefer \acrshortpl{ch} in closer proximity to the \acrshort{bs}.
Notably, \acrshort{leach-rlc} achieves a more balanced distribution of \acrshortpl{ch} for nodes with a $y$-coordinate greater than 50 m.
This balanced distribution is credited to \acrshort{leach-rlc}'s consideration of energy consumption by \acrshortpl{ch} during data transmission to the \acrshort{bs} and data reception from their cluster members—a feature absent in \acrshort{leach-c} and LEACH-CM.

Furthermore, \acrshort{leach-rlc}, \acrshort{leach-c} and LEACH-CM demonstrate higher \acrshort{pdr} than distributed protocols like \acrshort{leach}, EE-LEACH and LEACH-D (Fig.~\ref{fig:pdr}) thanks to the balanced energy distribution across the network.
Unlike distributed \acrshort{leach} protocols which eventually select nodes with low energy, causing \acrshortpl{ch} to deplete their energy within a round, resulting in a lower \acrshort{pdr}, \acrshort{leach-rlc}, \acrshort{leach-c} and LEACH-CM strategically manage energy consumption, leading to improved \acrshort{pdr} over traditional \acrshort{leach}.

Additionally, \acrshort{leach-rlc}'s superior performance can be linked to its lower average network energy consumption, as evident in Fig.~\ref{fig:average_energy_consumption}.
Notably, distributed \acrshort{leach} protocols exhibit multiple spikes in energy consumption, indicating failed attempts to select optimal \acrshortpl{ch} due to the inherent randomness of the protocol.
In this aspect, \acrshort{leach-rlc} consistently outperforms its counterparts, maintaining a lower average energy consumption throughout the simulation.
This is further corroborated in Fig.~\ref{fig:remaining_energy}, which depicts the remaining energy of the network over rounds.

An integral aspect in achieving a balanced distribution of energy consumption across the network lies in selecting the optimal number of \acrshortpl{ch}.
This key factor significantly influences the overall performance of clustering protocols.
Fig.~\ref{fig:histograms} visually presents the distribution of the number of \acrshortpl{ch} selected by each protocol.
Notably, distributed \acrshort{leach} protocols (Fig.~\ref{fig:leach_histogram}, Fig.~\ref{fig:ee_leach_histogram}, and Fig.~\ref{fig:leach_d_histogram}) exhibits a wide distribution, ranging from 0 to 14 \acrshortpl{ch}.
This broad spectrum is attributed to the inherent randomness of the protocol, leading to a non-uniform distribution of \acrshortpl{ch} across the network.
In contrast, \acrshort{leach-rlc}, \acrshort{leach-c} and LEACH-CM (Fig.~\ref{fig:leach_rlc_histogram}, Fig.~\ref{fig:leach_c_histogram}, and Fig.~\ref{fig:leach_cm_histogram}) strive to maintain the number of \acrshortpl{ch} at 10\% of the alive nodes.
However, \acrshort{leach-rlc} consistently achieves this target, while \acrshort{leach-c} and LEACH-CM show less frequency in maintaining this 10\% threshold and often present a notable number of \acrshortpl{ch} around 2.
This distinction highlights \acrshort{leach-rlc}'s superior ability to maintain a balanced and optimal number of \acrshortpl{ch} across the network.

\begin{figure*}[!htbp]
    \centering
    \subfloat[\label{fig:leach_new_ch_vs_energy_consumption}]{%
        \includegraphics[width=0.32\textwidth]{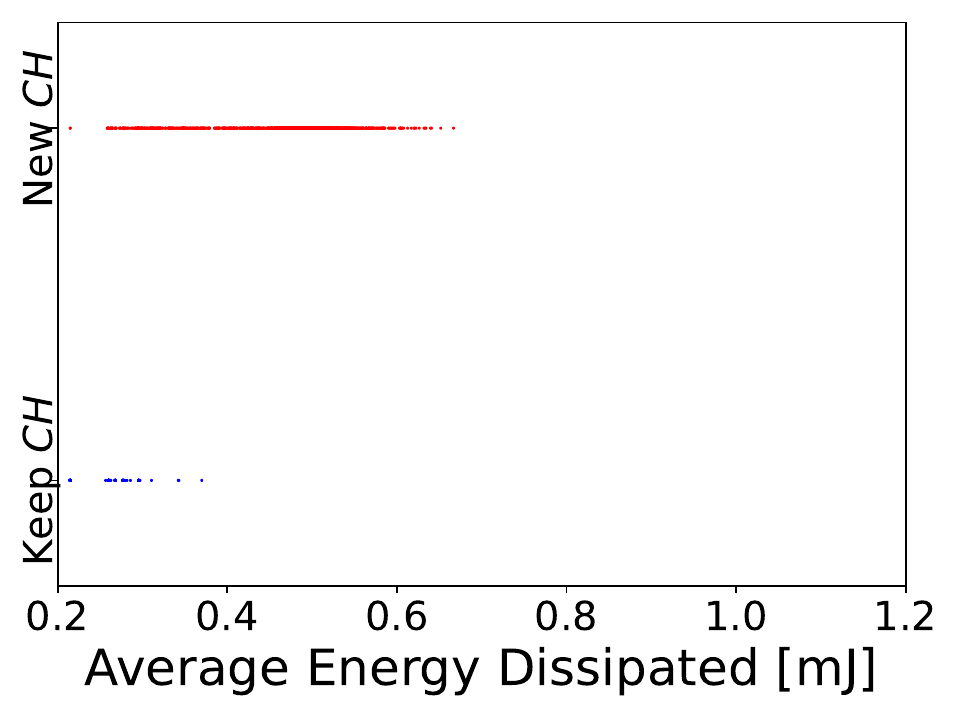}
    }
    \subfloat[\label{fig:leach_c_new_ch_vs_energy_consumption}]{%
        \includegraphics[width=0.32\textwidth]{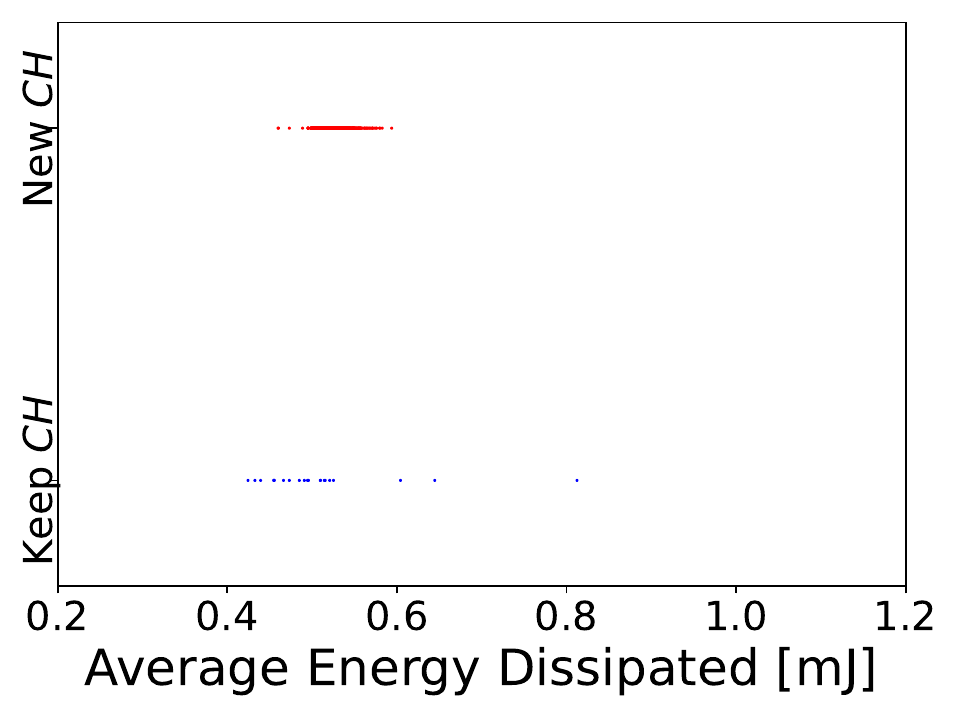}
    }
    \subfloat[\label{fig:leach_rlc_new_ch_vs_energy_consumption}]{%
        \includegraphics[width=0.32\textwidth]{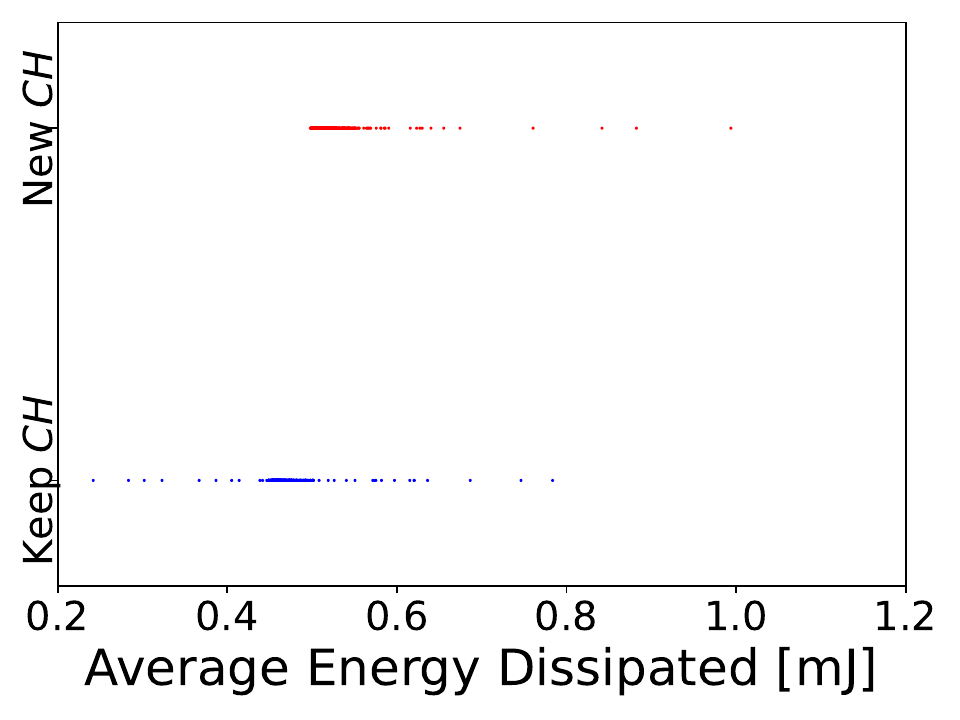}
    }
    \caption{Average network energy consumption versus the frequency of new cluster head selections. Panels (a), (b), and (c) illustrate the results for LEACH, LEACH-C, and LEACH-RLC, respectively.}
    \label{fig:new_ch_vs_energy_consumption}
\end{figure*}

\begin{figure*}[!htbp]
    \centering
    \subfloat[\label{fig:leach_new_ch_frequency}]{%
        \includegraphics[width=0.32\textwidth]{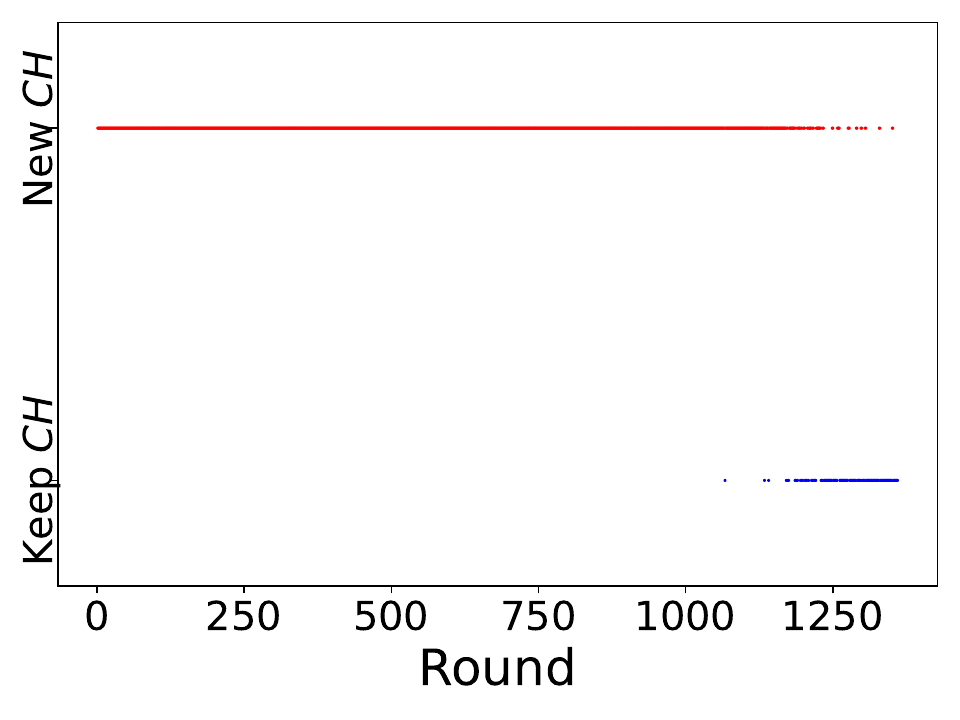}
    }
    \subfloat[\label{fig:leach_c_new_ch_frequency}]{%
        \includegraphics[width=0.32\textwidth]{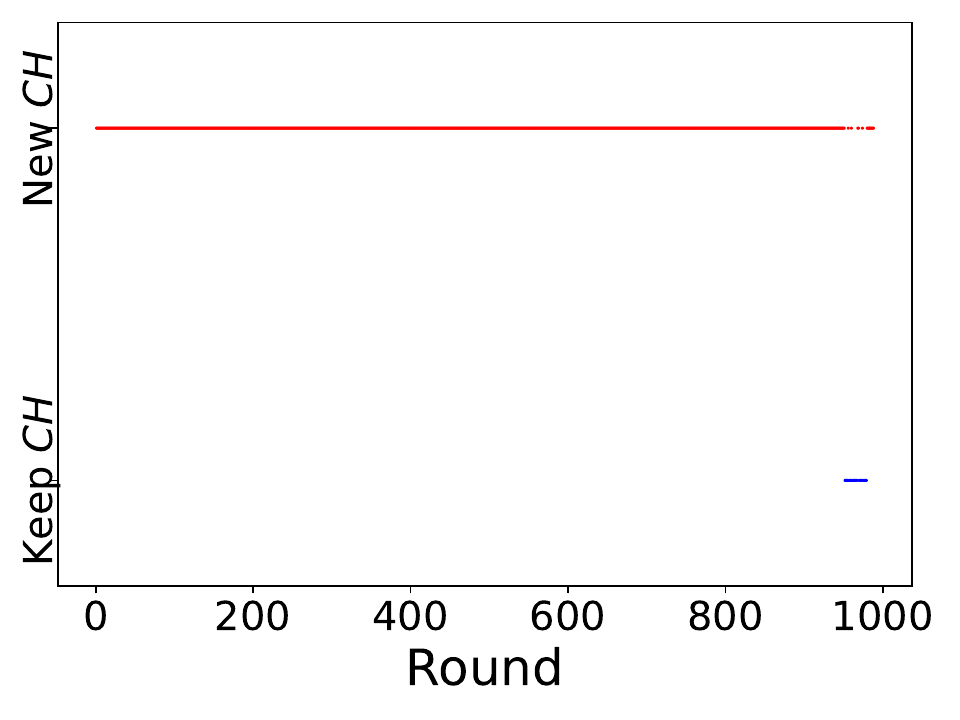}
    }
    \subfloat[\label{fig:leach_rlc_new_ch_frequency}]{%
        \includegraphics[width=0.32\textwidth]{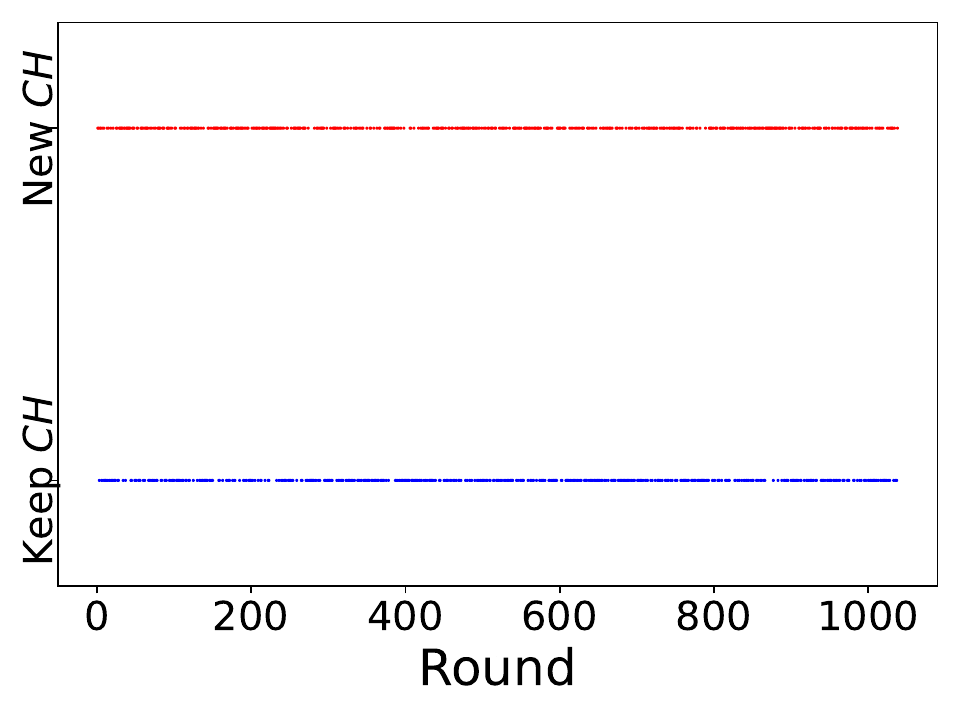}
    }
    \caption{Frequency of new cluster head selections across rounds. Panels (a), (b), and (c) present the results for LEACH, LEACH-C, and LEACH-RLC, respectively.}
    \label{fig:new_ch_frequency}
\end{figure*}

\subsection{Performance Analysis and Discussion}

Now, delving deeper into the performance analysis of \acrshort{leach-rlc}, we examine the correlation between network parameters and the number of selected \acrshortpl{ch} by the \acrshort{rl} agent.
Fig.~\ref{fig:correlation_matrix_rl} illustrates the correlation matrix for the \acrshort{rl} agent.
Notably, the action taken by the \acrshort{rl} agent exhibits a high correlation with the average energy dissipated by the nodes (\(\bar{E_d}\)), suggesting that the agent triggers a new action in response to changes in average energy dissipation.

This observation is corroborated in Fig.~\ref{fig:new_ch_vs_energy_consumption}, where we depict the average network energy consumption vs. the actions taken by the \acrshort{rl} agent (keeping the same set of \acrshortpl{ch} or generating new \acrshortpl{ch}).
Notably, \acrshort{leach} (Fig.~\ref{fig:leach_new_ch_vs_energy_consumption}) generates more new \acrshortpl{ch} than \acrshort{leach-c} (Fig.~\ref{fig:leach_c_new_ch_vs_energy_consumption}) and \acrshort{leach-rlc} (Fig.~\ref{fig:leach_rlc_new_ch_vs_energy_consumption}) due to its decentralized nature, neglecting the overall energy consumption of the network.
Conversely, \acrshort{leach-c} and \acrshort{leach-rlc} consider the network's overall energy consumption, with \acrshort{leach-rlc} showing a more uniform distribution of new \acrshortpl{ch} for higher values of average network energy consumption.

Fig.~\ref{fig:new_ch_frequency} further highlights the frequency of generating new \acrshortpl{ch} over rounds.
\acrshort{leach-rlc} (Fig.~\ref{fig:leach_rlc_new_ch_frequency}) exhibits a less frequent generation of new \acrshortpl{ch} compared to \acrshort{leach} (Fig.~\ref{fig:leach_new_ch_frequency}) and \acrshort{leach-c} (Fig.~\ref{fig:leach_c_new_ch_frequency}), reflecting the agent's adaptive evaluation of the network state at each round.
The agent weighs the performance of maintaining the current set of \acrshortpl{ch} against generating new \acrshortpl{ch}, effectively reducing control overhead, as evidenced in Fig.~\ref{fig:control_overhead}.
\acrshort{leach-rlc} consistently presents lower control overhead than its centralized counterparts, demonstrating the efficacy of the \acrshort{rl} agent in learning the optimal timing for control message transmission, a capability absent in \acrshort{leach-c} and LEACH-CM.
Although distributed \acrshort{leach} protocols exhibit zero control overhead, they do not guarantee an even distribution of \acrshortpl{ch} across the network, as demonstrated in previous figures.

\subsection {Research Questions Analysis}

The evaluation of \acrshort{leach-rlc} provides meaningful insights into the posed research questions.
Firstly, regarding \textit{RQ1}, our results showcase that \acrshort{leach-rlc} successfully reduces control overhead while maintaining or even enhancing network performance.
The intelligent clustering strategy, coupled with the adaptive \acrshort{rl} agent, ensures that the network operates efficiently without unnecessary overhead.

Moving on to \textit{RQ2}, the optimal frequency for generating new clustering solutions is a critical aspect of network management.
Our findings reveal that \acrshort{leach-rlc} dynamically adjusts the generation of new clusters based on the energy dynamics and network conditions.
This adaptability ensures an optimal balance, contributing to the extended network lifetime observed in our experiments.

Finally, addressing \textit{RQ3}, the opportune moment for triggering a new clustering solution is intricately linked to the energy consumption patterns.
\acrshort{leach-rlc} demonstrates a keen understanding of the network state, generating new clusters judiciously when energy consumption increases.
This strategic decision-making process significantly contributes to maintaining a balanced distribution of energy and prolonging the network's operational lifespan.

\subsection{Feasibility of the Proposed Approach in Real-World IoT Devices}

In real-world \acrshort{iot} deployments, sensor nodes typically perform fundamental tasks such as neighbor discovery, data packet generation, and communication with the \acrshort{ch} or sink.
Our approach introduces dynamic clustering through the use of control packets, which can be efficiently implemented using simple broadcast packets.
These packets distribute updated \acrshort{ch} and member information during each cluster reformation, ensuring that the nodes are always aware of their role and cluster structure.

For communication with the sink, nodes can use a \acrfull{mac} protocol such as \acrfull{tsch}, which provides high reliability and is well-suited for industrial \acrshort{iot} environments~\cite{hermetoSchedulingIEEE802154TSCH2017}.
Furthermore, these tasks—neighbor discovery, packet generation, cluster communication, and reliable data delivery—are typically already implemented in standard networked embedded systems.
These systems often operate with protocols like \acrfull{ipv6} over the \acrfull{6lowpan} layer, enabling seamless integration without imposing additional constraints or overhead~\cite{mulligan6LoWPANArchitecture2007}.

Additionally, constrained \acrshort{iot} devices are often capable of dynamically adapting their transmission power.
This feature can be leveraged to optimize energy consumption by adjusting the transmission power based on the distance to the intended destination, such as the \acrshort{ch} or sink.
For instance, nodes transmitting to a nearby \acrshort{ch} can reduce their power to conserve energy, while longer-range transmissions can increase power as needed.
This adaptability further supports the energy-efficient operation of the proposed clustering approach.

When a node is selected as a \acrshort{ch}, it must execute additional tasks, such as running an aggregation algorithm to process data collected from the cluster.
Depending on the application and the data of interest, simple aggregation functions, such as computing the average of the sensor data within a cluster, can be performed before forwarding the aggregated information to the sink.
This approach minimizes the communication overhead and reduces energy consumption across the network.

The only aspect that requires periodic updates is the dynamic reformation of clusters, specifically updating the \acrshort{ch} and member roles.
Nodes can utilize mechanisms like Orchestra, a \acrshort{tsch} scheduler, to automatically coordinate and establish communication with their respective \acrshortpl{ch}.
Since Orchestra manages slot and channel assignments autonomously, it further reduces overhead and ensures efficient communication within the network.

In summary, the proposed approach aligns with existing practices in networked embedded systems and leverages lightweight mechanisms that impose minimal additional requirements on the nodes.
By relying on broadcast packets for cluster management, well-established MAC protocols for communication, dynamic transmission power adjustment, and efficient data aggregation at \acrshortpl{ch}, our method can be seamlessly deployed in resource-constrained environments.

\section{Conclusion}
\label{sec:conclusion}

In this paper, we introduced \acrshort{leach-rlc}, an innovative clustering protocol designed to enhance the performance of \acrshort{wsn} applications.
Leveraging a \acrshort{milp}, \acrshort{leach-rlc} intelligently selects \acrshortpl{ch} and efficiently assigns nodes to clusters.
A key feature of \acrshort{leach-rlc} is the incorporation of a \acrshort{rl} agent, which significantly reduces control overhead by learning the optimal timing for generating new clusters.

The \acrshort{rl} agent in \acrshort{leach-rlc} employs a surrogate model, effectively estimating cluster heads and cluster members to expedite the learning process.
Our comprehensive evaluation of \acrshort{leach-rlc} delves into various performance metrics, including network lifetime, average energy consumption, control overhead, number of cluster heads, and the frequency of new cluster generation.

Results from the evaluation provide valuable insights into the optimal configuration of \acrshort{leach-rlc}, shedding light on the parameters that significantly influence its performance.
Particularly, the protocol's ability to balance energy consumption among nodes leads to a prolonged network lifetime.
This superior performance is attributed to \acrshort{leach-rlc}'s refined selection of \acrshortpl{ch} and their assignment to clusters.

Furthermore, our exploration into the correlation between network parameters and \acrshort{rl} agent actions unveils the agent's adaptability to changing energy dynamics.
The \acrshort{rl} agent demonstrates a nuanced response to the average energy dissipation of nodes, showcasing a balanced distribution of new \acrshortpl{ch} generation for varying energy consumption levels.

The answers to our research questions affirm that \acrshort{leach-rlc} effectively reduces control overhead without compromising network performance, determines the optimal frequency for generating new clusters, and strategically triggers new clustering solutions based on the network's energy dynamics.
These findings underscore the protocol's adaptability and efficiency in achieving prolonged network lifespan and energy balance.

While \acrshort{leach-rlc} demonstrates significant advancements, it is essential to acknowledge the limitations and potential areas for future research.
One limitation is its centralized approach, which may not scale efficiently in large networks.
Exploring decentralized RL approaches could enhance scalability and robustness.
Integrating \acrshort{leach-rlc} with edge computing paradigms could further reduce latency and improve real-time decision-making.

Future studies should also employ more realistic models that account for factors such as non-ideal wireless links, node mobility, and environmental interference, ensuring the framework's applicability in dynamic environments.
Additional research could refine \acrshort{leach-rlc}'s parameters, expanding its adaptability to diverse \acrshort{wsn} scenarios.
Systematic methods, such as grid search or hyperparameter optimization, could be used to analyze and optimize specific parameters, such as exploration rewards, thereby enhancing the framework's generalizability.

Although the training process focuses on fixed network topologies, network size and node placement significantly impact performance.
Sparse networks may benefit from reduced contention, leading to longer lifetimes, while denser networks face higher signaling overhead and energy consumption.
These factors influence convergence and overall network performance.
Future work could investigate scalability across various topologies and densities, providing deeper insights into the adaptability of \acrshort{leach-rlc} in different settings.

\bibliographystyle{IEEEtran}
\bibliography{references}

\begin{IEEEbiography}[{\includegraphics[width=1in,height=1.25in,clip,keepaspectratio]{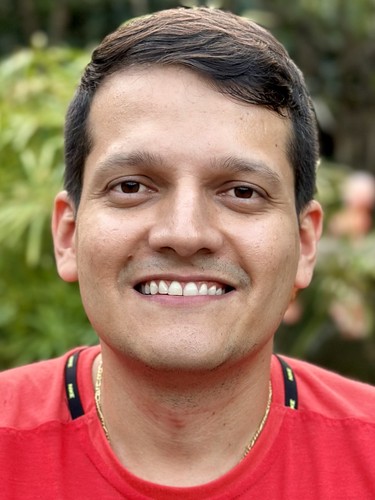}}]{F. Fernando Jurado-Lasso}
  (GS'18-M'21) received the Ph.D. degree in Engineering and the M.Eng. degree in Telecommunications Engineering both
  from The University of Melbourne, Melbourne, VIC, Australia, in 2020 and
  2015, respectively; a B.Eng. degree in Electronics Engineering in 2012 from
  the Universidad del Valle, Cali, Colombia. He is currently a postdoctoral
  researcher at the Embedded Systems Engineering (ESE) section of the
  Department of Applied Mathematics and Computer Science of the Technical
  University of Denmark (DTU Compute).

  His research interests include networked embedded systems, software-defined wireless sensor networks, machine learning, protocols, and applications
  for the Internet of Things.
\end{IEEEbiography}
\begin{IEEEbiography}[{\includegraphics[width=1in,height=1.25in,clip,keepaspectratio]{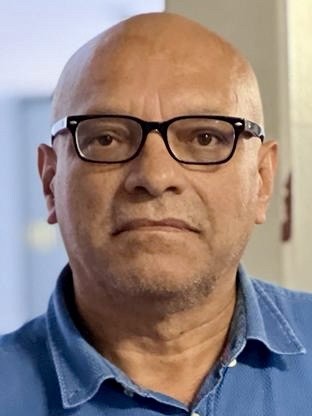}}]{J. F. Jurado}
  received the Doctorate and MSc degree in Physics both from Universidad del Valle, Cali, Colombia, in 2000 and 1986, respectively; he also holds a BSc degree in Physics from the Universidad de Nariño, Pasto, Colombia in 1984.
  He is currently a Professor with the Faculty of Engineering and Administration of the Department of Basic Science of The Universidad Nacional de Colombia Sede Palmira, Colombia. His research interests include nanomaterials, magnetic and ionic materials, nanoelectronics,  embedded systems, and the Internet of Things. He is a senior member of Minciencias in Colombia.
\end{IEEEbiography}
\begin{IEEEbiography}[{\includegraphics[width=1in,height=1.25in,clip,keepaspectratio]{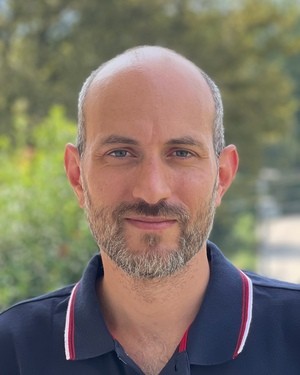}}]{Xenofon Fafoutis}
  (S'09-M'14-SM'20) received a PhD degree in Embedded Systems Engineering from the Technical University of Denmark in 2014; an MSc degree in Computer Science from the University of Crete (Greece) in 2010; and a BSc in Informatics and Telecommunications from the University of Athens (Greece) in 2007. He is currently a Professor with the Embedded Systems Engineering (ESE) section of the Department of Applied Mathematics and Computer Science of the Technical University of Denmark (DTU Compute). His research interests primarily lie in Wireless Embedded Systems as an enabling technology for Digital Health, Smart Cities, and the (Industrial) Internet of Things (IoT).
\end{IEEEbiography}

\balance

\vfill

\end{document}